\documentclass[9pt]{SelfArx} % Document font size and equations flushed left

\usepackage[utf8]{inputenc}
\usepackage[T1]{fontenc}
\usepackage{textcomp}
\usepackage{amsmath}
\usepackage{amssymb}
\usepackage{booktabs}
\usepackage{array}
\usepackage[switch]{lineno} % Line numbers for the review process
\usepackage[super,sort&compress]{natbib}
\setcitestyle{super,comma}

\newif\ifshownotes
\shownotesfalse
\definecolor{color1}{RGB}{0,0,90} % Color of the article title and sections
\definecolor{color2}{RGB}{0,20,20} % Color of the boxes behind the abstract and headings

\usepackage[unicode]{hyperref} % Required for hyperlinks
\usepackage{xurl}
\usepackage{epstopdf}
\usepackage{siunitx}
\usepackage[figurename=Fig.,justification=justified]{caption}
\DeclareCaptionLabelSeparator{nature}{\ \textbar\ }
\hypersetup{hidelinks,colorlinks,breaklinks=true,urlcolor=color2,citecolor=color1,linkcolor=color1,bookmarksopen=false,pdftitle={A hair-thin implant opens the entire depth of the mouse brain to revisiting the same structures at subcellular resolution},pdfauthor={Stiburek et al.}}

\ifshownotes
  \usepackage{pdfcomment}
  \newcommand{\note}[2]{\pdfcomment[author={#1},icon=Comment,color={1 0.8 0}]{#2}}
\else
  \newcommand{\note}[2]{}
\fi

\JournalInfo{~} % Journal information
\Archive{} % Additional notes (e.g. copyright, DOI, review/research article)

\PaperTitle{A hair-thin implant opens the entire depth of the mouse brain to revisiting the same structures at subcellular resolution}
\Authors{Miroslav Stib\r{u}rek\textsuperscript{1,*},
Tom\'{a}\v{s} Pik\'{a}lek\textsuperscript{1,*},
Petra Ondr\'{a}\v{c}kov\'{a}\textsuperscript{1,*},
Kristina Herz\'{a}nov\'{a}\textsuperscript{1},
Adrian Lorenz\textsuperscript{2},
Andr\'{e} Gomes\textsuperscript{2},
Julia Henschke\textsuperscript{3},
Endre Levente Marosi\textsuperscript{3},
Bhargavi Murthy\textsuperscript{3},
Petra Kolb\'{a}bkov\'{a}\textsuperscript{1},
Hongbo Jia\textsuperscript{3},
Sanja Bauer Mikulovic\textsuperscript{3,8},
Janelle Pakan\textsuperscript{3,6,7},
Alessio Attardo\textsuperscript{3,5,7},
Hana Uhl\'{i}\v{r}ov\'{a}\textsuperscript{1,$\dagger$}
and Tom\'{a}\v{s} \v{C}i\v{z}m\'{a}r\textsuperscript{1,2,4,$\dagger$}} % Authors
\affiliation{\textsuperscript{1}\textit{Institute of Scientific Instruments of CAS, Kr\'{a}lovopolsk\'{a} 147, 612 64 Brno, Czechia}}
\affiliation{\textsuperscript{2}\textit{Leibniz Institute of Photonic Technology, Albert-Einstein-Stra{\ss}e 9, 07745 Jena, Germany}}
\affiliation{\textsuperscript{3}\textit{Leibniz Institute for Neurobiology, Brenneckestra{\ss}e 6, 39118 Magdeburg, Germany}}
\affiliation{\textsuperscript{4}\textit{Institute of Applied Optics, Friedrich Schiller University Jena, Fr\"{o}belstieg 1, 07743 Jena, Germany}}
\affiliation{\textsuperscript{5}\textit{German Center for Mental Health (DZPG), Invalidenstraße 80, 10117 Berlin, Germany}}
\affiliation{\textsuperscript{6}\textit{Medical Faculty, Otto-von-Guericke University Magdeburg, Leipziger Stra{\ss}e 44, 39120 Magdeburg, Germany}}
\affiliation{\textsuperscript{7}\textit{Center for Behavioral Brain Sciences (CBBS), Otto-von-Guericke University Magdeburg, Universit\"{a}tsplatz 2, 39106 Magdeburg, Germany}}
\affiliation{\textsuperscript{8}\textit{Division of Neurophysiology, Center for Brain Research, Medical University of Vienna, Spitalgasse 4, 1090 Vienna, Austria}}
\affiliation{*\textbf{These authors contributed equally}}
\affiliation{$^\dagger$\textbf{Corresponding authors}: huhlirova@isibrno.cz, cizmart@isibrno.cz}

 % Keywords
\newcommand{\keywordname}{Keywords} % Defines the keywords heading name

\Abstract{\bf
Ultrathin multimode-fibre endoscopes provide minimally invasive access to subcellular structures throughout the depth of the mouse brain. Here we introduce a chronically implantable interface with a diameter of 144~\textmu m, about the thickness of a human hair, that extends this capability to longitudinal imaging, enabling repeated access to the same fine structures over extended periods. The system supports weeks-long tracking of individual dendritic spines, allowing quantitative comparison of their dynamics and survival across dendritic compartments. We further demonstrate longitudinal intracellular calcium imaging during spontaneous and stimulus-evoked activity in the same cells across days, as well as repeated measurements of blood flow in the same vessels across multiple brain regions. This chronic endomicroscopic platform provides stable, minimally invasive optical access for longitudinal subcellular-resolution imaging of brain structure, activity and blood flow across depth.
}
\begin{document}

\makeatletter
\rhead{\small\sffamily\bfseries Revisiting the same brain structures through a hair-thin implant\ --- \thepage/\pageref{LastPage}} % short running head
\makeatother

\fontsize{3.3mm}{4.3mm}\selectfont
\flushbottom % Makes all text pages the same height

\maketitle % Print the title and abstract box
% Notes on the title, authors and abstract (bubbles appear on page 1 when \shownotestrue)

%\linenumbers % Number lines for the review process

\thispagestyle{empty} % Removes page numbering from the first page

%----------------------------------------------------------------------------------------
%	ARTICLE CONTENTS
%----------------------------------------------------------------------------------------

% \section*{Introduction}  % Nature Methods: no heading for the introduction

Fluorescence microscopy has long been a cornerstone of preclinical neuroscience, enabling high-resolution imaging of neuronal structures within living tissues. Among these techniques, multiphoton microscopy has dominated \textit{in-vivo} applications by providing a powerful compromise between subcellular resolution and imaging depth\cite{Xu2024,Ji2016}. Using adaptive optics combined with an electrocardiogram-guided motion correction in 3-photon microscopy, individual spines can be resolved down to $\sim$\SI{0.9}{\milli\metre} and dendrites down to $\sim$\SI{1.4}{\milli\metre}\cite{Streich2021}. However, sub-micron microscopy beyond this depth in the brain tissue has not been achieved with non-invasive optical microscopy, owing to scattering and absorption of light.

To overcome these limitations, optical endoscopes have been employed, physically extending light delivery and collection to the target region. While gradient-index (GRIN) lenses are widely used\cite{Barretto2009,Barretto2011,Barretto2012,Bocarsly2015,Attardo2015,Kim2022}, recent advances in holographic endomicroscopy demonstrated high-quality imaging through multimode fibres of much smaller diameters. Notably, a 110-\textmu m wide holographic endomicroscope has been used to visualise neuronal structures across the entire depth of the mouse brain with sub-micrometre resolution\cite{Stiburek2023}. Its side-view design\cite{Silveira2021} enables imaging along the probe track during insertion, eliminating the need for invasive procedures such as tissue aspiration typically required for larger probes \cite{Mizrahi2004,Dombeck2010,Hazon2022}.

Holographic endomicroscopy dispenses with image-forming optics in the tissue altogether. Lens-based endoscopes relay an image through the probe, and both fundamental and practical (manufacturing) limits prevent such lenses from being made both long and narrow while preserving image quality\cite{Antonini2020,Sattin2024,Turtaev2025}. Holographic endomicroscopy does not rely on the waveguide to form an image. In a multimode fibre, mode mixing scrambles any input into a seemingly random speckle pattern, but because this transformation is linear and deterministic, it can be measured once as a transmission matrix (TM)\cite{Popoff2010,Cizmar2011} and then reversed: wavefront-shaping technology and methods tailor the light entering the fibre so that it leaves as a diffraction-limited focus, which is scanned point by point to form an image\cite{Cizmar2012,Turtaev2018,Ohayon2018,VasquezLopez2018}. All imperfections of the waveguide are thereby compensated at once. The probe is no thicker than the fibre, its resolution is set by the fibre's numerical aperture, and the same calibration places the focus numerically in different planes, yielding volumetric images without moving the probe\cite{Stiburek2023}.
% Previous version:
% Holographic endomicroscopy dispenses with image-forming optics in the tissue altogether. Lens-based endoscopes
% relay an image because a lens performs the optical Fourier transform that turns a plane wave into a focus;
% making the lens thinner shrinks its aperture and with it the resolution. A multimode fibre accepts light over
% a far larger numerical aperture for its diameter, but mode mixing scrambles any input into a seemingly random
% speckle pattern. Because this mixing is linear and deterministic, it can be measured once as a transmission
% matrix\cite{Popoff2010,Cizmar2011} and then reversed: wavefront-shaping technology and methods tailor the
% light entering the fibre so that it leaves as a diffraction-limited focus, which is scanned point by point to
% form an image\cite{Cizmar2012,Turtaev2018,Ohayon2018,VasquezLopez2018}. The probe is therefore no thicker than
% the fibre, its resolution is set by the fibre's numerical aperture rather than by its diameter, and the same
% calibration places the focus numerically in different planes, yielding volumetric images without moving the
% probe\cite{Stiburek2023}.

Despite their minimal diameter, insertion of such probes into delicate tissues like the brain can still cause microscopic damage to cells and vessels, disrupting the blood brain barrier and triggering inflammatory and wound-healing responses with each insertion\cite{Bocarsly2015,Karumbaiah2013,Fernandez2014}. These are characterised by increased activity of microglia and astrocytes, leading to local tissue remodelling over the course of hours to days. While such responses may not impede acute imaging, they preclude reliable chronic imaging of the same structures over time.

To address this limitation, we developed a minimally invasive implantable interface for chronic endomicroscopy. Our approach pairs an ultrathin glass capillary with a side-view multimode fibre (MMF) probe, increasing the diameter by only $\sim$30\%. This dimension remains well below the \SI{500}{\micro\metre} of the narrowest GRIN-lens-based systems in routine use\cite{Antonini2020,Sattin2024}. Similar to the guiding-tube preparations\cite{Bocarsly2015}, we implant the capillary during an initial surgical procedure and allow it to integrate into the tissue during recovery. Subsequently, we insert a side-view MMF probe into the capillary. This enables repeated, high-resolution imaging of the same cellular structures along nearly the entire length of the implant over periods of weeks. By keeping the imaging probe physically separated from the surrounding tissue, the approach avoids repeated probe--tissue contact and associated tissue perturbation.

Here, we applied the interface to chronic \textit{in-vivo} imaging of neurons acquiring 10 imaging sessions over 23 days. We resolved neuronal somata, axons with boutons and dendrites with spines along the accessible length of the implant. This enabled longitudinal tracking of neuronal structures across cortical and subcortical regions. The extended axial field of view allowed us to follow neuronal structures across distinct hippocampal subfields and quantify dendritic spine dynamics in basal and apical dendritic compartments within the same neuron. We further demonstrated repeated calcium imaging of the same neurons during spontaneous activity and in response to visual stimulation, as well as imaging of the vascular network and repeated measurements of blood-flow velocity in individual vessels at different depths. These demonstrations establish the versatility of the approach for longitudinal, depth-resolved imaging of neuronal structure, activity and vascular dynamics through a single ultrathin chronic interface.

\begin{figure}[tp] % single column; the caption must stay within ~12 lines (~125 words) to fit
   \includegraphics[width=\columnwidth]{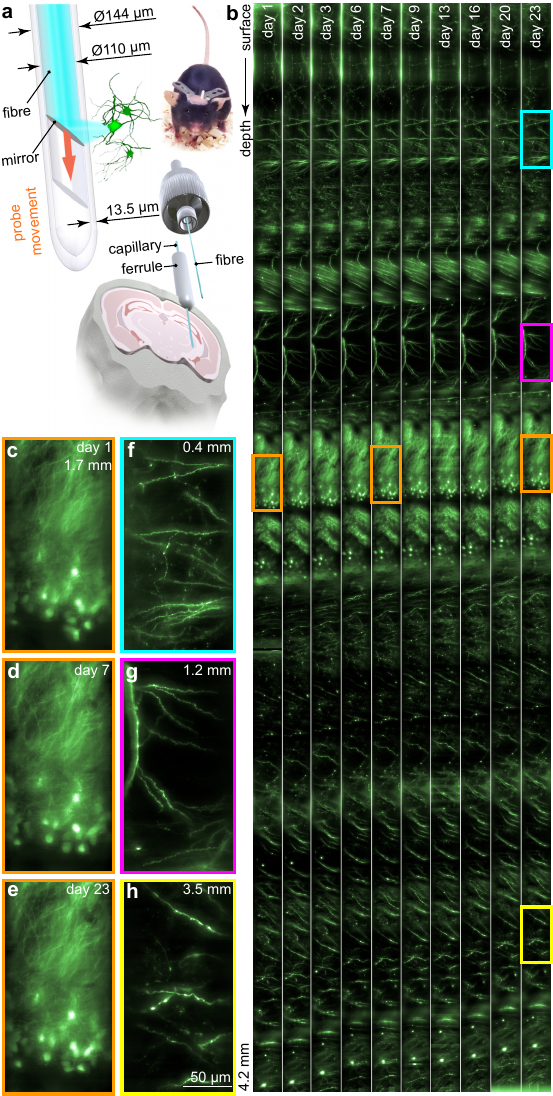}
   \caption{{\bf Endomicroscopic interface for chronic imaging.} {\bf a}, Capillary implant. Left, side-view multimode fibre (MMF) probe inside the glass capillary, imaging through its thin wall. Bottom right, implant and probe to scale with the mouse brain (model: smcinturff, Thingiverse, CC BY; adapted by the authors). Top right, mouse with the implant and a head plate. {\bf b}, Atlas view recorded while the probe was advanced along the capillary: maximum-intensity projection of 17 focal planes \SI{2.5}{\micro\metre} apart, brightness adjusted locally; 10 sessions over 23 days. {\bf c--e}, Neuronal somata and processes in the dentate gyrus at different time points. {\bf f--h}, Magnified regions of {\bf b}: cortical layer 5 (cyan, {\bf f}), hippocampal CA1 (magenta, {\bf g}) and hypothalamus (yellow, {\bf h}).}
   % Original (longer) caption -- does not fit under the figure in a single column:
   % {\bf Endomicroscopic interface for chronic imaging.} ({\bf a}) Capillary implant. Left, a side-view MMF probe
   % inserted into the glass capillary, enabling fluorescence imaging through the thin capillary wall. Bottom
   % right, the implanted capillary and MMF probe shown to scale relative to the mouse brain (mouse skull and brain
   % model source: smcinturff on Thingiverse; licensed under Creative Commons Attribution (CC BY); rendered and
   % adapted by the authors). Top right, a mouse bearing the implanted capillary and a head plate. ({\bf b})
   % Volumetric imaging record acquired as the endoscopic probe was advanced along the length of the capillary.
   % Seventeen focal planes were acquired at 2.5-\textmu m intervals and combined into a maximum-intensity
   % projection (MIP). The brightness was locally adjusted to account for dynamically varying fluorescence across
   % regions. The imaging procedure was carried out in 10 imaging sessions distributed over 23 days. ({\bf c--e})
   % Representative images of neuronal somata and processes in the dentate gyrus acquired at different time points
   % in the chronic imaging regime. ({\bf f--h}) Magnified regions of the atlas view in {\bf b} showing layer 5 of
   % the cortex (cyan, {\bf f}), hippocampal CA1 (magenta, {\bf g}) and hypothalamus (yellow, {\bf h}).
   \label{fig1}
\end{figure}

\section*{Results}

\subsection*{Endomicroscopic interface for high-resolution repeated access to deep brain structures}

% Previous wording: This enabled access to tissue depths of up to \SI{5}{\milli\metre}. The lateral spatial resolution varies across the field of view (FOV), with a median of \SI{1}{\micro\metre} and submicron resolution across $>$50\% of the FOV near the centre (see Supplementary Fig. 2).
We designed the endomicroscopic interface to satisfy the following criteria: (1) a minimal footprint for high tissue viability, (2) the ability to image at varying depths through a single implant while preserving homogeneous image quality, and (3) the capacity to repeatedly image identical regions of interest with high spatial precision. To this end, we developed the interface around a glass capillary with an outer diameter of \SI{144}{\micro\metre} and an inner diameter of \SI{117}{\micro\metre}. These dimensions were selected to minimise the implant footprint while providing sufficient mechanical stability (Fig. 1a). The implant length can be tailored to the experimental requirements by cleaving the capillary to the desired length. The distal end of the capillary is sealed during fabrication (see Online Methods and Supplementary Fig. 1), whereas the proximal end remains open and is secured within a ceramic ferrule. The ferrule is affixed to the skull with dental cement during implantation, sealing the surgical site and providing a stable, aseptic access point for probe insertion. A 110-\textmu m-diameter MMF (100-\textmu m core diameter, 0.37 NA) with a side-view termination is inserted into the capillary, allowing the surrounding tissue to be imaged through the capillary wall as the probe is translated along the implant. Imaging is possible along the active implant length corresponding to the full implant length except for the sealed distal segment, typically $\sim$\SI{250}{\micro\metre}. In this study, we imaged at depths of up to \SI{4.6}{\milli\metre}. The lateral spatial resolution varies across the field of view (FOV), with a median and mean of \SI{1.0}{\micro\metre} and submicron resolution across 52\% of the FOV, mainly near its centre (see Supplementary Fig. 2).

We tested the imaging performance in an anaesthetised head-fixed Thy1-GFP mouse capturing neuronal morphology from brain surface to depth of \SI{4.2}{\milli\metre} (Fig. 1b). Positional feedback from the stage drive mechanism enabled accurate stitching of individual images into an atlas view, sampling an area of over \SI{0.4}{\square\milli\metre}. We imaged 17 focal planes, \SI{2.5}{\micro\metre} apart using the numerical refocusing method\cite{Stiburek2023} and show their maximum-intensity projection. To optimise contrast and prevent signal saturation across the whole atlas view, we adaptively modulated the laser power in real time using feedback from the photomultiplier tube (PMT) signal. Additionally, we synchronised the scan sequence with the cardiac cycle in order to minimise motion artefacts (Online Methods, Atlas-view imaging).

\begin{figure*}[t]
   \includegraphics[width=\textwidth]{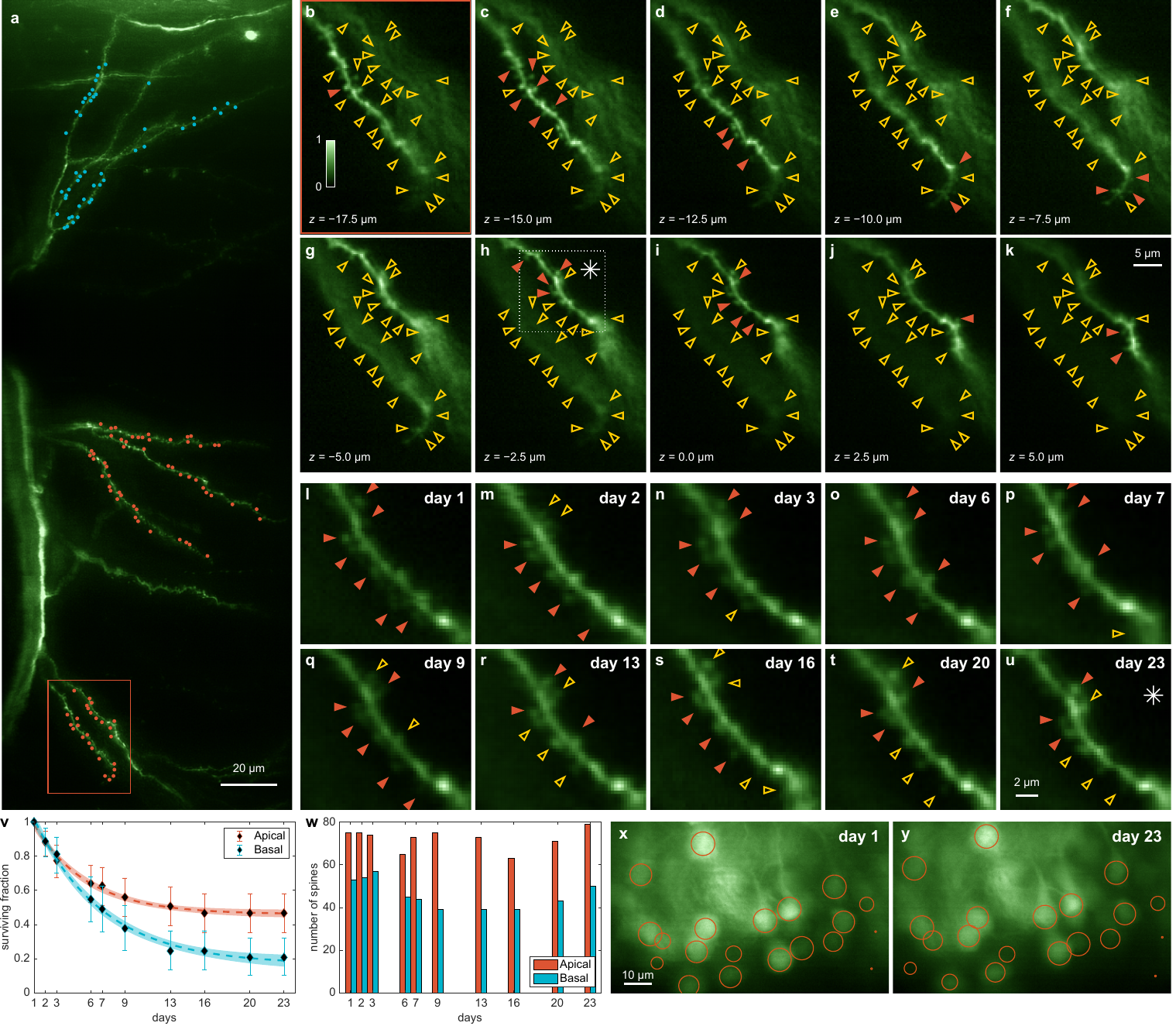}
   \caption{{\bf Longitudinal tracking of dendritic spine dynamics in the hippocampus.} {\bf a}, Maximum-intensity projection of segments of a CA1 pyramidal neuron (Fig.~\ref{fig1}g). Detected basal and apical dendritic spines from four basal and five apical dendrites are indicated in blue and red, respectively. {\bf b--k}, A magnified region of interest (ROI) of two apical dendrites shown in {\bf a}, imaged on day 23 (asterisk in {\bf u}). Ten focal planes are shown, with detected spines indicated by arrows. Red arrows mark spines detected in the displayed focal plane; yellow arrows mark spines detected in other focal planes. {\bf l--u}, Longitudinal imaging of the ROI indicated in {\bf h} by an asterisk recorded at indicated days across a period of 23 days. Red and yellow arrows mark spines within and outside the displayed focal plane, respectively. {\bf v}, Survival fraction of apical and basal dendritic spines relative to spine populations detected on day 1. Lines show exponential fits; shaded regions indicate the 95\% confidence intervals of the fits. Data were bootstrapped 10,000 times; error bars indicate 95\% confidence intervals. {\bf w}, Number of apical and basal dendritic spines detected at each day. {\bf x,y}, Single-focal-plane images of the dentate gyrus acquired on days 1 and 23, respectively. Somata detected in the displayed focal plane and tracked across adjacent focal planes are indicated by circles. Circle diameter represents the distance of each soma from the displayed focal plane, with smaller circles indicating greater distances (generated by the TrackMate plugin\cite{Ershov2022,Tinevez2017} of FIJI). All data were collected from a single mouse.}
   \label{fig2}
\end{figure*}

\begin{figure*}[t]
   \includegraphics[width=\textwidth]{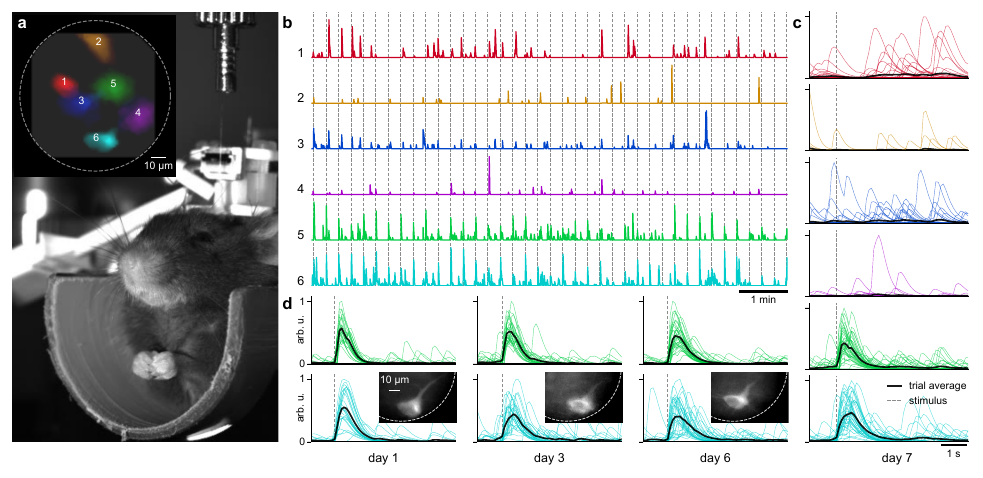}
   \caption{{\bf Chronic calcium imaging through the implanted capillary.} {\bf a}, Head-fixed mouse with the endomicroscopic probe inserted in the capillary implant and an example reference field of view (FOV) with neuronal regions of interest (ROIs) detected by the CaImAn software\cite{Giovannucci2019}. The circular region corresponds to the scanned FOV and the cropped square region marks the processed fraction (see Online Methods for details). {\bf b}, Normalised calcium traces from the neuronal ROIs in {\bf a} during visual stimulation. Stimuli are indicated with grey dashed lines. {\bf c}, Individual stimulus-evoked calcium responses (in colour) and their trial averages (in black). {\bf d}, Individual responses and corresponding trial averages for neurons 5 and 6 imaged longitudinally over four imaging sessions. Insets show cropped single-frame images of neuron 6 during the response. The edge of the circular FOV is indicated by a dashed line.}
   \label{fig3}
\end{figure*}

To evaluate the long-term performance, we repeated this atlas-view imaging in the same mouse over a three-week period (Fig. 1b). The tight fit between the probe and capillary (7-\textmu m clearance) allowed smooth, repeatable probe insertions while ensuring reliable return to the same regions of interest across sessions. Indeed, we could reliably re-identify identical neurons and their processes across 23 days, indicating high spatial precision and importantly, minimal optical or biological degradation over time (Fig. 1c-e).

The recorded images also maintained uniform resolution across imaging depths, as demonstrated by selected magnified views of neurons and their fine processes in the cortex, hippocampus and hypothalamus (Fig. 1f--h).

Notably, we successfully conducted the first imaging session immediately after the approved recovery period (7 days post-implantation)---significantly earlier than the standard two-week recovery window typically required for GRIN lens implants or cranial windows \cite{Barretto2011,Bocarsly2015,Attardo2015,Kilic2021,Park2019}. In contrast to these methods, we observed no signs of tissue opacity and minimal structural reorganisation. This early and stable imaging capability highlights the minimal invasiveness of the capillary interface and the readiness of the tissue for high-resolution optical interrogation shortly after surgery.

\subsection*{Longitudinal tracking of synaptic structural plasticity in the hippocampus}

Dendritic spine dynamics provide a sensitive readout of structural synaptic plasticity and are widely studied using longitudinal \textit{in-vivo} imaging\cite{Berry2017,Attardo2015,Mizrahi2004,CastelloWaldow2020,Chenani2022,Gu2014}. Several studies have shown that dendritic spine density and dynamics differ between the cortex and hippocampus, highlighting regional differences in structural synaptic plasticity\cite{Attardo2015,Gu2014,Bloss2018,Holtmaat2006}. These observations underscore the need for approaches that provide subcellular access to multiple brain regions and enable their longitudinal comparison within the same animal.

We therefore tested whether our implant could support chronic monitoring of dendritic spines. As a proof of principle, we chose the hippocampal circuit as a representative test bed and, in a Thy1-GFP mouse, targeted a CA1 pyramidal neuron with dendrites extending across the stratum pyramidale, stratum radiatum and stratum lacunosum-moleculare, while simultaneously visualising neurons in the dentate gyrus (Fig. 2a--u, x--y). Using numerical refocusing of volumetric recordings, we identified and tracked dendritic spines on its basal and apical segments (Fig. 2a) across 10 sessions spanning 23 days (Fig. 2l--u, Supplementary Video 1). On average, 119 dendritic spines across five apical and four basal dendrites were tracked per session. To quantify spine stability, we computed the established survival fraction (SF), defined as the fraction of spines present at day 1 that remained detectable at each subsequent time point\cite{Attardo2015}.

Consistent with previous reports, the SF exhibited an overall exponential decay over time (Fig. 2v). In this representative dataset, apical dendritic spines exhibited a slower decline of SF compared to basal dendritic spines. Exponential modelling indicated that apical spine survival approached an asymptotic value of approximately 46\% within the imaging period, whereas basal spine survival continued to decrease throughout the 23-day observation window. Extrapolation of the fitted decay suggested an asymptotic survival fraction of approximately 17\%, which would be reached beyond the duration of the experiment. Total spine density, reflecting the combined effects of spine elimination and formation, is shown in Fig. 2w.

In parallel, all 21 dentate gyrus granule cell somata within the field of view remained stably identifiable across the full imaging period (Fig. 2x-y, Supplementary Video 2), highlighting the robustness of long-term registration across hippocampal subfields.

Together, these results demonstrate that chronic holographic endomicroscopy enables stable, long-term volumetric imaging across the intact hippocampal circuit, providing sufficient spatial and temporal resolution to identify differences in synaptic dynamics across distinct dendritic compartments.

\subsection*{Chronic imaging of calcium dynamics}

In acute experiments, mechanical disturbance during endoscopic probe insertion leads to increases in intracellular calcium of nearby neurons. Indeed, neuronal activity has been shown to display changes in firing patterns sometimes sustained over tens of minutes before they recover\cite{Fiath2019}. This transient disruption poses a major limitation for endoscopic imaging approaches that rely on relative movement between the probe and tissue to access new fields of view, resulting in prolonged waiting times and inefficient data collection. In contrast, our chronic implant provides a stable, pre-recovered interface between the probe and the brain, enabling immediate imaging without inducing artefactual neuronal activation.

We evaluated the performance of this approach in an awake Thy1-GCaMP6f (GP5.17) mouse, recording calcium activity during visual stimulation (Fig.~3). As expected, relative motion between the probe and the tissue did not evoke neuronal activity: the animal moved during imaging, as seen in the simultaneous camera recordings (Supplementary Videos 4 and 5) and in the drift of the field of view relative to the tissue, which we corrected during processing, but no neuronal activity correlated with these movements. Visual stimuli elicited reliable responses in some neurons in the visual cortex, particularly pyramidal cells of layer V (Fig.~3b,c, Supplementary Videos 3 and 4). Neurons lying below cortex showed rather spontaneous activity both in somata and their processes (Supplementary Video 5). Importantly, individual active neurons could be reliably tracked across sessions, demonstrating precise spatial registration but also sustained tissue viability (Fig.~3d).

\subsection*{Chronic vascular imaging and single-vessel blood flow measurement}

Acute insertion of an endoscopic probe can transiently perturb the local vasculature and so compromise measurements of blood flow. Mechanical disruption of small vessels can cause extravasation of red blood cells or leakage of intravascular fluorescent agents, and tissue compression can alter local haemodynamics \cite{Jung2004, Wang2024}. Chronic implantation separates the insertion from the imaging and allows the vasculature to recover before measurements are made. Consistent with this, we observed no signs of vascular injury or dye leakage at the start of imaging.

Through the implant we mapped the vasculature, labelled with fluorescein-conjugated dextran, along the entire active length (Fig. 4a), and selected vessel segments at depths from the cortex down to the subthalamic region. The same eight segments were re-identified in all seven imaging sessions over 19 days (Fig. 4b), indicating that the interface is mechanically and optically stable. In each session, the focus was scanned repeatedly along the segment (cyan in Fig. 4b), yielding a kymograph in which moving blood cells appear as inclined streaks (Fig. 4c). After correction for uneven illumination and for the non-uniform spacing of the scan points along the vessel (Fig. 4d), the flow velocity follows from the inclination of the streaks. We estimated it by shearing the kymograph and finding, at each time point, the shear at which the streaks become vertical (Fig. 4e,f; see Online Methods). This gives the velocity at the line-scan rate, \SIrange{0.8}{3.8}{\kilo\hertz}, over records of about \SI{3}{\minute} (Fig. 4g,h).

\begin{figure*}[htb]
   \includegraphics[width=\textwidth]{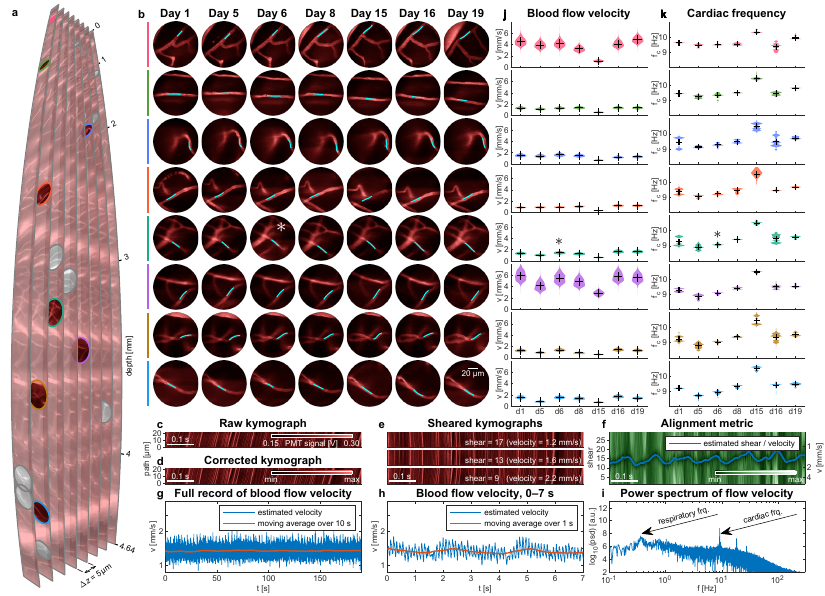}
   \caption{{\bf Longitudinal measurement of blood flow velocity in individual vessels.} {\bf a}, Atlas view of the labelled vasculature along the implant. The fields of view used for velocity measurements are highlighted; those shown in {\bf b} are colour-coded. {\bf b}, Reference images of eight fields of view (rows) in seven sessions (columns), with the scan path (cyan) on the same vessel segment each time. {\bf c--i}, Velocity estimation for one record (asterisks in {\bf b}, {\bf j} and {\bf k}). {\bf c}, Raw kymograph along the scan path. {\bf d}, Kymograph after correction and resampling. {\bf e}, The kymograph in {\bf d} sheared by three candidate values (time shift across the scan path, in line periods); at the correct shear, the streaks of moving blood cells become vertical. {\bf f}, Alignment metric against time and shear, with the estimated shear and the corresponding velocity. {\bf g}, Velocity over the full record. {\bf h}, First \SI{7}{\second} of {\bf g}. {\bf i}, Power spectrum of {\bf g}; arrows mark the respiratory and cardiac frequencies. {\bf j}, Velocity in each vessel (colours as in {\bf b}) across sessions; violins, distribution over the record; crosses, mean. {\bf k}, Cardiac frequency from the spectral peak (crosses); shapes show the profile of the peak, stretched tenfold in frequency for visibility.}
   \label{fig4}
\end{figure*}

Mean velocities ranged from 0.8 to \SI{6.0}{\milli\metre\per\second} between vessels and varied by 6--22\% between sessions for a given vessel (Fig. 4j). The power spectra of the velocity showed a sharp peak at the cardiac frequency, \SIrange{8.7}{9.9}{\hertz}, and a second peak at the respiratory frequency, \SIrange{0.4}{0.7}{\hertz} (Fig. 4i). The cardiac frequency could thus be read from the flow itself in every record (Fig. 4k). On day 15, the velocity in all eight vessels was reduced to between a quarter and just over half of its value in the other sessions, while the cardiac frequency was higher than in any other session (\SIrange{10.3}{10.5}{\hertz}). The protocol and the anaesthesia were unchanged, and outside the imaging session the animal behaved normally, as on any other day. We attribute this to the physiological state of the animal on that day; its consistent detection across vessels and depths illustrates the sensitivity of the measurement to systemic changes.

Together, these measurements show that the chronically implanted interface enables repeated quantitative assessment of blood flow in the same vessels across brain depths and imaging sessions.

\section*{Discussion}

We present a chronically implantable optical interface that enables repeated high-resolution imaging of deep brain regions using an ultra-thin capillary combined with a side-view MMF-based holographic endomicroscope. The design builds on previous guiding-tube concepts while substantially reducing the implant footprint. Compared with conventional GRIN lens-based approaches, the displaced brain volume is reduced by more than an order of magnitude, which may improve tissue preservation and recovery\cite{Karumbaiah2013}. In agreement with this, stable and high-quality optical access was observed from the first imaging session, 7 days after implantation.

The transverse imaging geometry provides several advantages of prism-based approaches\cite{Andermann2013,Chia2009,Redman2022} while avoiding their main limitations. Microprism implants require substantial tissue displacement and can restrict implantation depth and optical performance. In contrast, our approach enables imaging along the entire active length of the implant rather than from a fixed focal plane at the distal tip. This provides flexible access to inter- and intralaminar structures across a broad depth range while maintaining a minimal implant size.

The rotational symmetry of the capillary interface further expands the accessible imaging volume. The side-view MMF probe (or the sample) may potentially be rotated around its axis to enable circumferential imaging (Supplementary Video 6). Combined with axial translation, this approach would enable spiral scanning strategies that could extend the sampled volume into the mesoscale regime, with projected fields of view exceeding \SI{3}{\square\milli\metre}.

We demonstrated the versatility of chronic holographic endomicroscopy across three applications: longitudinal dendritic spine imaging, stimulus-evoked calcium imaging and single-vessel blood flow velocity measurements. The imaging periods reported here were set by the experimental protocol, not by the implant. In separate animals, the implants were tolerated for more than six months without complications, and we are not aware of any mechanism that would prevent imaging over such periods; establishing the practical limit beyond the periods reported here will, however, require further investigation.

% Previous version:
% Each modality was demonstrated in a single animal, in line with the 3Rs principle of reduction. The device-level measurements (resolution, repeatable probe positioning and re-identification of the same structures across sessions) do not depend on the individual animal, whereas the biological observations, such as the difference in survival between apical and basal spines, are presented as illustrations of what the method can measure rather than as general findings.
The protocol reported here is the outcome of several iterations of development, each prompted by a failure identified \textit{in vivo}. For example, filling the implant with deionised water trapped air bubbles, which migrated between the probe and the capillary wall and caused severe aberrations; filling with ethanol, whose lower surface tension prevents bubble formation, eliminated them. The development stages, the experiments that identified and resolved each issue and the animals involved are summarised in Supplementary Note 1 and Supplementary Table 1. All data shown were acquired with the final protocol.

% Previous version: With the final protocol, each modality is shown in one animal, in line with the 3Rs principle of reduction. The blood flow measurements were repeated in a second animal over three imaging sessions with consistent results. Structural and calcium imaging were performed in 3 and 2 animals, respectively, with comparable outcomes.\note{Tomas}{Placeholders: numbers of animals (and imaging sessions) with the final protocol for structural and calcium imaging; confirm that the second blood-flow animal agrees with the first, and decide whether to show it (e.g. a Supplementary Fig.) -- reviewers will ask to see the replicate. This paragraph adds about 120 words to the main text.} 
In line with the 3Rs principle of reduction, we stopped the experiments with each modality as soon as its functionality had been confirmed beyond reasonable doubt. The blood flow measurements were repeated in a second animal over three imaging sessions with consistent results, and structural and calcium imaging were performed in three and two animals, respectively, with comparable outcomes; the figures show one animal per modality. The device-level measurements (resolution, repeatable probe positioning and re-identification of the same structures across sessions) do not depend on the individual animal, whereas the biological observations, such as the difference in survival between apical and basal spines, are presented as illustrations of what the method can measure rather than as general findings.

We tracked individual dendritic spines in dorsal CA1 pyramidal neurons over a period of more than three weeks while simultaneously imaging dentate gyrus neurons during the same session. The image quality enabled quantification of spine survival fractions and, to our knowledge, provided the first longitudinal comparison of spine dynamics between basal and apical dendritic compartments of the same neuron. Apical spines showed higher persistence than basal spines, consistent with differences in their afferent inputs and local inhibitory circuits. The measured survival fractions followed the expected exponential decay but were lower than those reported using higher-resolution imaging approaches\cite{Attardo2015,Redman2022}. This difference likely reflects the lower spatial resolution of single-photon endoscopy, manual annotation variability and potentially the earlier onset of imaging after cranial surgery, when inflammatory responses are known to influence spine turnover\cite{Xu2007}. Therefore, these measurements should be interpreted as demonstrating relative differences in spine dynamics rather than providing precise estimates of absolute turnover rates. Nevertheless, they demonstrate that chronic holographic endomicroscopy provides sufficient resolution and stability to track structural plasticity across hippocampal subfields and subcellular compartments \textit{in vivo}.

We next demonstrated longitudinal calcium imaging across multiple brain regions. Stimulus-evoked calcium responses could be repeatedly measured in the same neurons across imaging sessions, although the limited field of view restricts the number of neurons that can be recorded simultaneously. Larger neuronal populations can be accessed by sequentially repositioning the fibre probe within the implanted guide, trading simultaneous recording for broader spatial coverage. This flexibility enables repeated functional measurements across distributed brain regions while retaining the minimally invasive nature of the chronic interface.

Finally, we demonstrated longitudinal measurements of blood flow in the same vessels over time and across brain depths. The ability to repeatedly access individual vascular segments provides a means to monitor vascular dynamics longitudinally in regions that are difficult to access with conventional optical approaches. Combined with the neuronal imaging capabilities of the system, this approach could enable longitudinal investigation of the relationships between vascular dynamics, neuronal activity and structural plasticity across brain regions.

Together, these capabilities establish chronic holographic endomicroscopy as a minimally invasive platform for longitudinal microscopic imaging of brain structure and function. The combination of cellular- and subcellular-resolution imaging, extended depth access and long-term stability enables repeated interrogation of neuronal and vascular processes across brain regions through a single ultrathin chronic interface. This provides a methodological framework for investigating how neuronal activity, synaptic structure and vascular dynamics evolve over extended timescales in health and disease. Although demonstrated here in mice, the approach could be adapted to other animal models.

Looking ahead, recent advancements in bending-resilient graded-index fibres\cite{Turtaev2025} could further extend MMF-based holographic endomicroscopy beyond head-fixed experiments. Combined with the chronic side-view architecture presented here, these fibres may enable cellular-resolution imaging of deep brain circuits during natural behaviour, learning and disease progression.

% \section*{Methods}\label{sec:mm}
% Methods are available in the online version of the paper.  (added by the journal at publication)

\section*{Acknowledgments}
We would like to acknowledge support from the European Innovation Council (NEUROGATE, 101213906), the European Research Council (724530, 101069245, 101082088 and 101158010), the Ministry of Education, Youth and Sports of the Czech Republic (CZ.02.1.01/0.0/0.0/15\_003/0000476, CZ.02.1.01/0.0/0.0/16\_013/0001775 and Czech-BioImaging, LM2018129), the Grant Agency of the Czech Republic (GA23-07432S and 25-18280S), the European Union's H2020-RIA (101016787), Czech Technology Agency -- Centre of Advanced Electron and Photonic Optics (TN02000020) and the Institute of Scientific Instruments of the CAS. We acknowledge the Czech-BioImaging facility of the Institute of Scientific Instruments CAS, ISI-MR LM2023050 for hosting our animal experiments. We would like to express our gratitude to Zenon Star\v{c}uk jr.\ and his team for sharing their facilities, assistance with animal models and advice. We thank Ond\v{r}ej Nov\'{a}k for discussions about chronic surgery procedures and \c{C}a\u{g}atay Ayd\i{}n for his advice on long-term chronic implantation.

\subsection*{Authors contributions:}
M. S.: conceptualisation, methodology, investigation, software, validation, visualisation, writing -- review \& editing;
T. P.: conceptualisation, methodology, investigation, software, validation, visualisation, writing -- review \& editing;
P. O.: conceptualisation, methodology, investigation, validation;
K. H.: formal analysis, visualisation, writing -- review \& editing;
A. L.: methodology, resources;
A. G.: formal analysis, data curation, visualisation, writing -- review \& editing;
J. H.: methodology, resources;
E. L. M.: methodology, resources;
B. M.: formal analysis, data curation;
P. K.: investigation;
H. J.: conceptualisation, resources, writing -- review \& editing;
S. B. M.: conceptualisation, resources, writing -- review \& editing;
J. P.: conceptualisation, resources, writing -- review \& editing;
A. A.: conceptualisation, resources, supervision, formal analysis, writing -- review \& editing;
H. U.: conceptualisation, funding acquisition, methodology, investigation, data curation, formal analysis, supervision, project administration, visualisation, writing -- original draft, writing -- review \& editing;
T. \v{C}.: conceptualisation, funding acquisition, resources, supervision, project administration, methodology, software, investigation, formal analysis, visualisation, writing -- review \& editing.

\subsection*{Competing interests:}
M. S., T. P., P. O., H. U., A. L. and T. \v{C}. are inventors on a related patent application: PCT/EP2026/079259 (Apparatus and method for chronic imaging through multimode fibers).

\subsection*{Data, code, and materials availability:}
During the peer-review process, the codes, the original (raw) and the processed data necessary to reproduce our results are available from \url{https://isibrnocz.sharepoint.com/:f:/s/comphot/IgCn4-KKT61xR4OXPuYq7yr6AR8snYHqx0TnkYOdBxMrZUY?e=6Pxdej}. Following the peer-review, the materials will be stored at zenodo.org. Data acquired during the development of the method (Supplementary Note 1) are not part of the deposited dataset because of their volume (several terabytes) and are available from the corresponding authors on reasonable request.

\bibliographystyle{nature}
\bibliography{refs}

\section*{Online Methods}\label{sec:om}

\subsection*{Implant fabrication}

% Previous short version:
% The capillaries for implants were drawn from a silica preform tube (Heraeus; outer and inner diameters of \SI{22}{\milli\metre} and \SI{19}{\milli\metre}) on a fibre-drawing tower at the Leibniz Institute of Photonic Technology to an outer diameter of \SI{144}{\micro\metre} and a wall thickness of \SI{13.5}{\micro\metre}, with a protective acrylate coating. Segments with a total length of 11--\SI{12}{\milli\metre} and an implantable length of 4.0--\SI{5.3}{\milli\metre}, depending on the targeted brain region, were cleaved and sealed at one end with a plasma arc, forming a rounded tip (Supplementary Fig. 1) that shortens the active imaging length by approximately \SI{250}{\micro\metre}. The implants were then bonded into ceramic ferrules (Thorlabs CFLC230) using UV-curable optical adhesive (Norland NOA61). The drawing process is described in Supplementary Methods 2.
The capillaries for implants were drawn from a capillary fibre. The drawing was performed at the Leibniz Institute of Photonic Technology (Leibniz IPHT). A Heraeus silica preform tube with an outer diameter of \SI{22}{\milli\metre} and an inner diameter of \SI{19}{\milli\metre} was vertically fed into the furnace at a rate of \SI{0.7}{\milli\metre\per\minute} using a linear translation stage. The preform was heated to \SI{2050}{\degreeCelsius} in a resistively heated Astro HR80 furnace. At this temperature, the viscosity of the silica decreases sufficiently to enable thermal elongation during the drawing process.

The capillary fibre was pulled using the main capstan of the drawing tower. For mechanical protection and handling, an acrylate coating (DSM 3471-3-14) was applied using a split-die open-cup applicator. The coating material adheres to the fibre surface and was subsequently cured by UV irradiation using a Heraeus Noblelight UV lamp system.

The capillary fibre was drawn at a speed of \SI{12}{\metre\per\minute} with a drawing tension of \SI{0.49}{\newton} (\SI{50}{gf}). To compensate for the collapse of the capillary caused by surface tension, an overpressure of \SI{80}{\pascal} was applied to the preform using a Mensor CPC4000 pressure controller. The outer diameter was measured inline using a Zumbach ODAC14XY diameter measurement system. The fibre was drawn to the target outer diameter using a feedback loop that adjusted the drawing tower’s capstan speed. During continuous drawing, samples were collected directly from the process spool, and their inner diameter was measured offline using a Keyence VHX-4000 digital microscope. This procedure yielded a capillary with an outer diameter of \SI{144}{\micro\metre} and a wall thickness of \SI{13.5}{\micro\metre}.

The thin-walled capillaries were thermally sealed at one end using a controllable plasma arc generated by a large-diameter splicing system (LDS 2.5, 3SAE), resulting in a rounded tip (Supplementary Fig. 1). This reduced the active imaging length of the implant by approximately \SI{250}{\micro\metre}. The capillaries were then assembled under a stereoscopic microscope into ceramic ferrules (Thorlabs CFLC230) using a UV-curable optical adhesive (NOA61, Norland). The implantable length was typically set to 4--\SI{5}{\milli\metre}. The entrance ends of the capillaries were cleaved using a ruby fibre scribe (S90R, Thorlabs), resulting in a total implant length ranging from 11--\SI{12}{\milli\metre}.

\subsection*{Fibre probe fabrication}

Side-view fibre probes were fabricated from CeramOptec Optran WFGe multimode fibre (core/cladding diameter of 100/\SI{110}{\micro\metre}, numerical aperture of 0.37). The fibre was stripped of its acrylate coating and cleaved at a right angle (Fujikura CT-105) to a length of \SI{45}{\milli\metre}. The fibre was then secured in an ST/PC connector using a UV-curable adhesive (NOA 65, Norland), with the input facet aligned with the end of the connector's ceramic ferrule. The distal end of the fibre was subsequently polished at \SI{45}{\degree} using a custom-built fibre polishing system. An aluminium mirror coating was deposited onto the polished facet by vacuum evaporation (JEOL JEE-420).

\subsection*{Animals}

All animal experiments were conducted in accordance with protocols approved by the Branch Commission for Animal Welfare of the Ministry of Agriculture of the Czech Republic (permissions no. 29-2023-P for chronic and 47/2020 and 49/2020 for acute experiments). All transgenic lines were obtained from Jackson Laboratory, Maine, USA. The STOCK Tg(Thy1-EGFP)MJrs/J (Thy1-GFP line M, Stock No: 007788) was used in structural imaging experiments and the C57BL/6J-Tg(Thy1-GCaMP6f)GP5.17Dkim/J (GP5.17 mouse line, Stock No: 025393) was used for calcium imaging. Wild-type C57BL/6J (Envigo) were used in experiments involving blood flow measurements. Both male and female mice aged 8--32 weeks were used. In separate animals kept under the same protocol, the implants were tolerated for more than six months without complications. % Previous version: In total, {[}N{]} mice were implanted, of which {[}N{]} were imaged in this study; {[}reasons for exclusion, if any{]}.\note{Tomas}{Hana: please fill in the numbers of implanted and imaged animals (ARRIVE guidelines), and the reasons for any exclusions.}
In total, 23 mice were used: two in acute pilot experiments, 14 for the development of chronic implantation (Supplementary Note 1, Supplementary Table 1) and seven with the final protocol. Data from one animal per modality are shown in the figures; all data acquired with the final protocol are available (see Data, code, and materials availability). 
%Of the latter, three were imaged for the data shown (one for structural imaging, one for calcium imaging and two for blood flow measurements); {[}reasons for exclusion, if any{]}.\note{Tomas}{Hana, Tomas: fill in the numbers (ARRIVE guidelines); they must add up with Supplementary Table 1. State how many final-protocol implants yielded usable imaging -- reviewers will ask for this success rate.} 
Mice were housed in a \SI{12}{\hour} light /\SI{12}{\hour} dark cycle at \SI{21}{\degreeCelsius} to \SI{23}{\degreeCelsius} and 50--60\% relative humidity with \textit{ad libitum} access to food and water.

\subsection*{Surgical procedures}

Mice were anaesthetised with isoflurane (Vetflurane, Cymedica, Czech Republic) delivered in 100\% oxygen (3.5--5\% for induction; 1.5--2.0\% for maintenance). Buprenorphine (\SI{0.04}{\milli\litre}, \SI{0.05}{\milli\gram\per\kilo\gram}) and dexamethasone (\SI{0.13}{\milli\litre}, \SI{2}{\milli\gram\per\kilo\gram}) were administered intraperitoneally to provide analgesia and prevent brain oedema, respectively. Subcutaneous injections of atropine (\SI{0.06}{\milli\litre}, \SI{0.05}{\milli\gram\per\kilo\gram}) and glucose (5\%, \SI{0.1}{\milli\litre}) were given to facilitate respiration and maintain hydration, respectively.

The mouse was placed in a stereotaxic frame on a feedback-controlled heating pad to maintain body temperature at \SI{37}{\degreeCelsius}. After fur removal and local application of lidocaine (2\%, \SI{0.05}{\milli\litre}), the scalp over the target region was excised via a circular incision. The edges of the skin were affixed with tissue adhesive (3M Vetbond Veterinary Tissue GLue). The skull was cleaned with sterile cotton swabs and a scalpel, then lightly scored in a crosshatch pattern to improve adhesion. A thin layer of cyanoacrylate glue (Loctite) was applied to the exposed bone. Coordinates of the target site were determined using the mouse brain atlas\cite{Franklin2019} and marked relative to bregma.

A custom-designed head plate, held in a mounting holder pre-aligned with the insertion stage (see Alignment of implant, animal and probe), was affixed over the target area using cyanoacrylate gel (Loctite) and further secured with dental cement (Jet Denture Repair Powder and Liquid, Lang Dental). For post-operative analgesia, ketoprofen (\SI{0.06}{\milli\litre}, \SI{5}{\milli\gram\per\kilo\gram}) was administered subcutaneously at least one hour prior to the end of the procedure. The mouse was then removed from the ear bars and stabilised by the head plate.

A high-speed drill with a \SI{0.3}{\milli\metre} tip was used to create a $\sim$\SI{2}{\milli\metre} diameter craniotomy centred over the target region. A small incision was made in the dura to minimise the tissue compression during capillary insertion. The capillary, mounted on a 3D manipulator (Unit Mini 25, Luigs \& Neumann), was lowered vertically at $\sim$\SI{10}{\micro\metre\per\second} until the base of the ferrule contacted the brain surface. The surrounding skull was dried, and the implant was initially stabilised using a small amount of UV-curable cement (Optibond FL, Kerr). All exposed bone and the head plate were then reinforced with dental cement (Jet Denture Repair Powder and Liquid, Lang Dental). A protective plastic cap (CAPL, Thorlabs) was placed over the implant tip.

Anaesthesia was discontinued, and the mouse was transferred to a heated recovery chamber. Once fully ambulatory, the mouse was returned to its home cage, which was free of obstructions that could interfere with the implanted device. Postoperative analgesia was maintained with subcutaneous injections of ketoprofen (\SI{0.06}{\milli\litre}, \SI{5}{\milli\gram\per\kilo\gram}) administered daily for two days following surgery.

\subsection*{Alignment of implant, animal and probe}

Imaging along the full length of the implant requires axial alignment of the probe with the capillary and the same probe geometry during calibration and imaging (Supplementary Note 1, Stage 3). During head plate attachment, a mounting holder aligned with the insertion stage held the plate perpendicular to the insertion axis. After attachment, the plate was secured in the holder bars for craniotomy and implantation. Two USB cameras (12-MP Sony IMX708 sensor, fixed focus; B03041, Arducam) were aligned with the three-axis manipulator (Unit Mini 25, Luigs \& Neumann; Supplementary Fig. 4) and provided a reference for aligning the capillary with the insertion axis before implantation. For imaging, the head plate was secured in the same holder bars on the endomicroscope stage to reproduce the implantation geometry. Fibre probes were secured in ferrules with bore diameters closely matched to the fibre diameter and mounted in an ST/PC connector (a49196, Alternativo). During calibration, the probe was inserted into a capillary secured in a groove milled into the calibration basin frame, maintaining the same straight geometry as during imaging (see Optical setup, calibration and filling of the implant; Supplementary Fig. 3).

\subsection*{Optical setup, calibration and filling of the implant}

The optical setup (Supplementary Fig. 3) was adapted from a previously reported design\cite{Stiburek2023} and extended for chronic imaging through capillary implants. Light from a \SI{488}{\nano\metre} continuous-wave laser was shaped by a temperature-stabilised digital micromirror device (DMD)\cite{Rudolf2021}, whose holograms control two orthogonal polarisations independently, and relayed onto the proximal facet of the side-view MMF probe. Fluorescence collected by the fibre was separated from the excitation light by a dichroic mirror and detected by the PMT.

% Previous short version of the calibration paragraph:
% Before each imaging session, the TM was measured by phase-shifting interferometry, coupling approximately 21,000 plane waves into the fibre and recording the output on a camera. To reproduce the optical path \textit{in vivo}, the fibre was calibrated inside a water-filled glass capillary of a calibration basin, in which light passes through the fibre, the capillary wall and water, as it does through the implant into tissue (Supplementary Methods 3).
The optical power of the 488-nm collimated continuous-wave laser was controlled using a motorised half-wave plate (H1) and polarising beam splitter (PBS1). A second half-wave plate--polarising beam splitter pair (H2, PBS2) controlled the power ratio between the signal and reference beams. In the signal arm, lenses L3 and L4 magnified the laser beam to overfill the DMD chip. The DMD was illuminated at an incidence angle of approximately \SI{24}{\degree} and operated at a pattern refresh rate of \SI{30303}{\hertz}. A beam sampler (BSA) diverted a fraction of the illumination beam to a photodiode (PD) to monitor the optical power. The temperature of the DMD was stabilised using a Peltier-based unit\cite{Rudolf2021}.

The hologram displayed on the DMD contained two carrier frequencies generating two spatially separated first-order beams in the DMD's Fourier plane. One of the beams passed through a 0.5-inch half-wave plate (H5) rotating its polarisation by \SI{90}{\degree}. These two beams are thus orthogonally polarised and both polarisations could be controlled separately. The beams were subsequently recombined using a 40-mm calcite beam displacer (BD). An additional half-wave plate (H6) and quarter-wave plate (Q) were used to set linear polarisation before coupling into the fibre. A dichroic mirror (DM) in the path reflected the illumination towards the fibre. Two telescopes formed by lenses L6, L7 and L8, L10 relayed and demagnified the DMD plane onto the proximal fibre facet, with undesired diffraction orders eliminated along the optical path.

The light was then propagated through the fibre, exiting sideways at the distal tip. An objective (OBJ) and a tube lens (L11) magnified the output from the fibre onto the camera (CAM1) where it interfered with the reference beam.

The reference beam passed through a shutter (S) and a pair of lenses (L1 and L2) expanded the beam to fill the camera chip. A non-polarising beam splitter (BS) combined the reference and signal beams.

During sample imaging, the emitted fluorescence was collected within the numerical aperture of the fibre and propagated to the proximal end. It was spectrally filtered by the DM and a barrier filter (F). Lens L9 focused the signal onto the PMT.

\emph{Calibration.}

% Previous wording: Filling capillaries with water during calibration and also during imaging decreased the gradient of the refractive indices from the glass fibre to the surrounding environment, thereby reducing total internal reflection at the edges of the fibre facet. This prevented a reduction of the field of view.
Before each imaging session, the setup was calibrated by acquiring the TM using phase-shifting interferometry. Approximately 21,000 plane waves with different transverse wavevectors were sequentially coupled into the fibre as input modes for TM acquisition. The TM was measured at full camera resolution over a selected rectangular region and numerically propagated to the required imaging planes. Output sampling points were then selected from these propagated TMs, and the corresponding illumination patterns were calculated. The spatial extent and sampling density of the selected points depended on the imaging modality (atlas, blood-flow or calcium imaging), as described in the corresponding sections below. To eliminate optical aberrations during imaging, the calibration was carried out along an optical path mimicking the one \textit{in vivo}. To this end we built a calibration basin (CB in Supplementary Fig. 3) -- a small tank enclosed with a cover glass on each side, with capillaries glued into grooves milled in its metal frame. During the calibration, the tank was filled with water. The capillaries had both ends open and were therefore filled with water via capillary action. For the calibration, the fibre was inserted into one of the capillaries. Filling the capillary with liquid, water during calibration and ethanol during imaging (Supplementary Note 1, Stage 4), decreased the refractive-index contrast between the glass fibre and its surroundings, thereby reducing total internal reflection at the edges of the fibre facet. This prevented a reduction of the field of view. The two liquids differ only slightly in refractive index (1.33 for water and 1.36 for ethanol), and the difference affects only the thin liquid layer between the fibre and the capillary wall; its effect on imaging is negligible. Similar to the \textit{in-vivo} environment, the light during the calibration passed from the fibre, through the capillary and through water in the basin, mimicking the surrounding tissue. All components of the calibration unit were mounted on a motorised XYZ stage (components marked with an asterisk in Supplementary Fig. 3), which was retracted after TM acquisition to allow positioning of the animal. The calibration basin was mounted on a separate manual XYZ stage (marked with a double asterisk in Supplementary Fig. 3).

\emph{Filling the implant before imaging.}

The stage was then lowered to withdraw the fibre from the calibration capillary and driven to the filling position, where a smaller-diameter glass capillary connected to a syringe was aligned above the implant. Under visual guidance from CAM2 and CAM3, the filling capillary was inserted into the implant and advanced to its bottom by raising the stage. Ethanol was used as the immersion medium because its low surface tension reduces bubble formation within the implant. The implant was filled from bottom to top by applying pressure to the syringe while continuously lowering the stage. After filling, the stage was driven to the imaging position, and the imaging fibre was aligned above the implant. The fibre was then inserted into the implant, and imaging was initiated.

\subsection*{Animal preparation for imaging}

Before each atlas-view imaging session, anaesthesia was induced with 3\% isoflurane in oxygen and maintained at 1.5\% for the rest of the experiment. The mouse was positioned on a platform over a heating blanket and secured via its head plate. A pulse oximeter sensor was attached to the shaved thigh to monitor heart rate. The platform was mounted on the endomicroscope's three-axis motorised XYZ stage. After removal of the protective implant cap, the stage was driven to the calibration position. Two navigation cameras (CAM2 and CAM3 in Supplementary Fig. 3) provided visual guidance for precise alignment of the imaging fibre above the calibration capillary. The stage was then raised along the Z axis to insert the fibre into the capillary. The fibre output was imaged using the calibration camera (CAM1 in Supplementary Fig. 3), and the TM was measured.

The implant was then filled with ethanol, whose low surface tension prevents the formation of bubbles, through a thinner glass capillary inserted to the bottom of the implant, after which the imaging fibre was inserted and imaging was started (see Optical setup, calibration and filling of the implant).

For blood-flow imaging, the animal was anaesthetised and placed on a heating blanket. Following topical anaesthesia of the eye with oxybuprocaine hydrochloride eye drops (\SI{4}{\milli\gram\per\milli\litre}), \SIrange{0.05}{0.1}{\milli\litre} of fluorescein isothiocyanate--dextran (2 MDa, FD2000S, Sigma-Aldrich; 5\% in PBS) was administered by retro-orbital injection. The mouse was subsequently head-fixed to the platform and positioned in the endomicroscope. The calibration, implant filling and imaging procedures were then performed as described above.

For calcium imaging, animals were habituated to head fixation using progressively increasing durations of \SIrange{5}{60}{\minute}. Before each imaging session, the mouse was head-fixed to the platform, and the body was restrained in a tube to minimise movement and stress. The calibration and implant-filling procedures were performed before imaging.

At the end of each imaging session, the fibre probe was withdrawn from the capillary, and the implant was fitted with a protective cap. Anaesthesia was discontinued, and the animal was removed from the platform and transferred to a heated cage for recovery. Heating was discontinued once the animal had fully regained consciousness.

\subsection*{Atlas-view imaging}

Atlas-view images were acquired by continuously capturing side-view images during the insertion of the fibre probe into the capillary at a constant velocity. The individual frames were subsequently stitched during post-processing to generate the final composite images.

The TM of the fibre was measured at a plane approximately \SI{25}{\micro\metre} from the outer wall of the capillary. This TM was then numerically propagated to a series of $\pm 8$ parallel planes in \SI{2.5}{\micro\metre} increments, resulting in a total of 17 imaging planes spanning working distances from approximately \SI{5}{\micro\metre} to \SI{45}{\micro\metre}. Focused excitation points within any of these planes were generated by displaying corresponding patterns on the DMD. During probe insertion, each plane was imaged sequentially; however, the random-access capability of the DMD allowed for flexible scanning order once the required pattern sequences were loaded into memory.

% Previous short version:
% To minimise motion artefacts, acquisition was synchronised with the heartbeat. The heart rate was measured with a pulse oximeter (MouseOx Plus, STARR Life Sciences) before imaging and remained stable during the session. Each plane was imaged in rectangular blocks (typically $400 \times 8$ pixels at a pitch of \SI{0.286}{\micro\metre}), each acquired within one cardiac cycle, and the blocks of all 17 planes were acquired in sequence while the probe advanced at approximately \SI{1}{\micro\metre\per\second}, so that neighbouring pixels, including those in neighbouring planes, were recorded at the same cardiac phase. The excitation power was adjusted continuously to the local fluorescence and recorded for normalisation, and the blocks were stitched using the stage coordinates, without image registration (Supplementary Methods 4).
\emph{Cardiac-synchronised acquisition.} To minimise motion artefacts induced by cardiac activity, image acquisition was synchronised with the animal's heartbeat. Approximately 15 minutes after anaesthesia induction---allowing sufficient time for animal mounting, endomicroscope calibration, and capillary filling---the heart rate was measured using a pulse oximeter (MouseOx Plus, STARR Life Sciences). As the heart rate remained stable throughout the imaging session, this initial measurement was used to define the entire scanning sequence.

Imaging within each plane was performed in rectangular blocks, with block dimensions (width and height) optimised to maximise the width of the field of view while ensuring that each block was acquired exactly within a single cardiac cycle. Typically, blocks were $\sim$400 pixels ($\sim$\SI{105}{\micro\metre}) wide and 8 pixels ($\sim$\SI{2}{\micro\metre}) tall, though these values varied slightly with heart rate. Pixels within each block were scanned column-wise at maximum resolution corresponding to pixel pitch of \SI{0.286}{\micro\metre}.

Blocks from all 17 planes were acquired in sequence, resulting in a complete imaging cycle spanning exactly 17 heartbeats. This ensured that spatially adjacent pixels, including those in neighbouring planes, were imaged at the same phase of the cardiac cycle, effectively reducing motion-related artefacts. This approach compensated only for cardiac motion; other sources, such as respiration, were not corrected.

The insertion speed of the fibre probe was chosen so that the vertical displacement of the stage over one full imaging cycle matched the block height minus one pixel (allowing a 1 px overlap for stitching). This typically corresponded to an insertion velocity of approximately \SI{1}{\micro\metre\per\second}.

To account for spatial variability in fluorescence expression across brain regions, the excitation laser power was continuously adjusted during imaging, ensuring consistent utilisation of the detector's dynamic range. Laser power was monitored via a photodiode and used to normalise the acquired fluorescence images during post-processing.

In post-processing, dark frame subtraction was followed by normalisation of the PMT signal using the recorded excitation power. The rectangular imaging blocks were then stitched into a continuous atlas-view. Given the mechanical stability provided by the capillary interface and the absence of tissue deformation during fibre insertion, block positions could be directly inferred from stage coordinates, eliminating the need for image registration. In the 1 px overlap regions, pixel intensities were averaged to ensure seamless transitions between adjacent blocks.

\subsection*{Identification, tracking, and data processing of dendritic spines and dentate gyrus granule cell somata}

The dendritic spines were tracked on the atlas-view volumetric records described before using ImageJ in combination with the TrackMate plugin\cite{Ershov2022,Tinevez2017}. Each spine was marked at the focal plane where it exhibited its maximum diameter and assigned a unique identifier. Every identifier comprises the 3-dimensional spatial location of the spine (x,y, and z) along with the day of observation and the spine type (apical or basal). When a spine was not observed on a given day (i.e., the synaptic site was unoccupied) and later reappeared in a subsequent imaging session at the same apparent location (now occupied synaptic site), it was assigned a new unique identifier and considered a new spine. Based on the identified spines, the survival fraction for each type (apical and basal) was calculated as the number of spines persisting from day 1 to subsequent days, normalised by the total number of spines present on day 1 (Fig. 2v).

% Previous short version:
% Confidence intervals of the survival fraction were estimated by bootstrap resampling (10,000 resamples), and the survival fraction of each spine type was fitted with an exponential decay with a plateau (Supplementary Methods 5). The image in Fig. 2a is a maximum-intensity projection after blind 3D deconvolution, and the images in Fig. 2l--u were registered to that of day 1 by normalised cross-correlation (Supplementary Methods 5).
Since only a single dataset was collected (i.e., longitudinal tracking of spines over days in a single mouse model), a bootstrap resampling approach was employed to estimate the confidence intervals for the fraction of surviving spines. The observed data was resampled 10,000 times, with replacement, and without assuming a specific distribution. From the 10,000 bootstrap resamples, the 95\% confidence interval was computed as the 2.5\textsuperscript{th} and 97.5\textsuperscript{th} percentiles of the resulting distributions and represented as error bars in Fig. 2v.

The fraction of surviving spines for each type was fitted using least-squares regression with an exponential decay curve:

\[\mathrm{SF} = a\,e^{-bt} + c,\]

where \(\mathrm{SF}\) is the survival fraction \(a\) represents the amplitude above the plateau, \(b\) is the decay rate constant, \(t\) is time and \(c\) is the function plateau, which represents the long-term persistent fraction of stable spines that do not decay over the measured time window. Note that the constraint enforcing a survival fraction of 1 at the starting day was not imposed, but left as free parameter during the fitting procedure.

The fits yielded \(a_\mathrm{apical} = 0.65 \pm 0.05\); \(b_\mathrm{apical} = 0.21 \pm 0.04\,\mathrm{day}^{-1}\) and \(c_\mathrm{apical} = 0.46 \pm 0.03\) for the apical spines and \(a_\mathrm{basal} = 1.01 \pm 0.06\); \(b_\mathrm{basal} = 0.17 \pm 0.03\,\mathrm{day}^{-1}\) and \(c_\mathrm{basal} = 0.17 \pm 0.05\) for the basal spines.

The confidence intervals for the fitted curves (also represented in Fig. 2v) were computed using linearised prediction intervals based on the Jacobian of the fitted model and the residual variance.

The atlas-view in Fig. 2a was obtained by performing 3D deconvolution of the volumetric stack followed by maximum-intensity projection. In this case 10 iterations of blind deconvolution were employed, since the point spread function (PSF) of the holographic endomicroscope may vary slightly with scanning position, particularly across different focal planes, and because the true 3D PSF is not available for each scanned position. The PSF for deconvolution was therefore assumed to be a gaussian beam. The PSF lateral standard deviation \((\sigma_{xy})\) was defined from the diffraction-limited lateral resolution \(\left( FWHM = \frac{\lambda_{0}}{2NA} \right)\), where \(\lambda_{0}\) is the wavelength in vacuum, \(NA\) is the fibre numerial aperture, and FWHM is the full width at half maximum, which relates to the lateral standard deviation as: \(FWHM = 2\sqrt{2\ln{(2)}}\sigma_{xy}\). Similarly, the PSF axial standard deviation \(\left( \sigma_{z} \right)\) was defined from the Rayleigh range \(\left( z_{r} \right)\) as: \(z_{r} = \frac{n\lambda_{0}}{\pi{NA}^{2}},\ \)where \(n\) is the refractive index of the medium (here assumed close to water: 1.333), which is related to the axial standard deviation as: \(z_{r} = 2\sqrt{2\ln{(2)}}\sigma_{z}\).

Compensation of small displacement between images of different days in Fig. 2l-u was performed using intensity-based registration via normalised cross-correlation. The translation parameters of each image were obtained using the image of day 1 (Fig. 2l) as reference and applied to each of them respectively.

Similarly to the dendritic spines, dentate gyrus granule cells were tracked across the different imaging sessions from the measured volumetric time series using ImageJ in combination with the TrackMate plugin.

\subsection*{Calcium imaging and image processing}

Focal excitation points were calculated from the measured TM within a circular field of view approximately \SI{105}{\micro\metre} in diameter. The focal-point grid was initially decimated by a factor of two in both lateral dimensions, resulting in a pixel pitch of \SI{0.572}{\micro\metre}. Nine focal planes, spanning \SI{-20}{\micro\metre} to \SI{+20}{\micro\metre} relative to the calibrated plane in \SI{5}{\micro\metre} increments (corresponding to working distances of approximately \SIrange{5}{45}{\micro\metre}), were used. The resulting focal-point patterns were loaded into the memory of the DMD. Within each plane, focal points were scanned in a raster-like pattern.

The calibrated grid was used to acquire high-resolution reference images at a frame rate of approximately \SI{1.1}{\hertz}. For neuronal activity imaging, the grid was further decimated by a factor of three in both lateral dimensions, resulting in a pixel pitch of \SI{1.716}{\micro\metre} and a frame rate of approximately \SI{10}{\hertz}, sufficient to resolve the dynamics of GCaMP6f fluorescence.

During imaging, the fibre was slowly inserted into the implant while continuously acquiring images, allowing the desired FOV to be located. Once the target structures were identified and brought into focus, fibre movement was stopped. Visual stimulation was then initiated using a blue LED positioned adjacent to the mouse, and neuronal activity was recorded using the reduced-resolution scanning pattern. The driving voltage of the stimulation LED was sampled simultaneously with the fluorescence signal to provide precise temporal synchronisation between visual stimulation and fluorescence recordings. At the end of each FOV, a high-resolution reference image was acquired using the calibrated focal-point grid.

Throughout almost all experiments, the mouse was illuminated with an infrared LED (\SI{850}{\nano\metre}) and imaged using two cameras (Basler a2A1920-160umBAS, Supplementary Video 4 and 5) to monitor the animal and synchronise the camera images with the endomicroscopic recordings. At the end of the imaging session, the fibre was retracted from the implant.

The acquired imaging data were processed using the open-source CaImAn software package\cite{Giovannucci2019}. Image sequences were first imported, and rigid motion correction was performed on the largest square region inscribed within the circular field of view. This region was used for registration because the original circular field of view contains image boundaries within the frame, which can make shift estimation unreliable. For each frame, CaImAn estimated a single two-dimensional translational shift relative to the registration template. The resulting frame-wise shifts were subsequently applied to the corresponding full-field images, thereby preserving the complete circular field of view for downstream analysis.

Neuronal activity was extracted using the constrained non-negative matrix factorisation algorithm for microendoscopic one-photon imaging (CNMF-E)\cite{Zhou2018}. Putative neuronal sources were initialised based on the local correlation and peak-to-noise ratio of the fluorescence signal (Corr-PNR). Temporal components were estimated using OASIS (Active Set method for Automatic Inference of Spikes)-based constrained deconvolution. Following model fitting and automated component evaluation, candidate neurons were further inspected manually to remove false-positive detections, which predominantly originated from the image boundaries.

The fluorescence traces shown in Fig. 3b correspond to the denoised temporal components estimated by CNMF-E for the accepted neuronal sources. These traces were obtained from the CNMF-E temporal activity matrix and represent the inferred calcium dynamics after separation of the neuronal signal from background activity and noise during model fitting. The denoised traces were used for visualisation and subsequent stimulus-response analysis. For trial-averaged responses (Fig. 3c-d), the traces were aligned to stimulus onset, determined from the rising edges of the simultaneously recorded LED signal, and averaged across repeated stimulation trials.

\subsection*{Vascular imaging and blood flow velocity estimation}

To visualise the vasculature throughout the depth of the implant, atlas-view imaging was performed before the first and after the last blood-flow session; the latter, acquired on day 20, is shown in Fig. 4a. The TM was measured in a plane approximately \SI{25}{\micro\metre} from the nearest point on the capillary's outer cylindrical surface. The TM was numerically propagated to $\pm 8$ parallel planes in \SI{5}{\micro\metre} increments yielding 17 planes in total. Cardiac synchronisation was omitted for this application, as it did not significantly improve image quality for large vascular structures. Instead, each scan cycle covered a single horizontal line of 416 pixels (\SI{119}{\micro\metre}) in each of the 17 planes in turn, and as the probe advanced, the next cycle scanned the line below. The scanned points thus formed a horizontal plane, perpendicular to the probe, that swept down through the tissue together with the probe.
% Previous wording: Instead, each plane was imaged using a 416 px (119 um) wide by 1 px tall block, effectively generating horizontal cross-sectional views of the sample in each cycle.

Based on the resulting vascular map, blood vessel segments were selected for flow measurements. In each imaging session, the fibre was re-inserted to the pre-defined depths of the selected segments. Full-field images ($173 \times 173$ foci at a pitch of \SI{0.572}{\micro\metre}, corresponding to every second pixel of the calibration camera; field of view \SI{99}{\micro\metre} in diameter) were acquired, and the selected vessel was brought into focus using precomputed DMD patterns. Reference images were obtained by averaging multiple frames (Fig. 4b).

\emph{Line scans.} For each vessel, 8--37 scan points were placed manually along the visible vessel segment (scan path length \SIrange{16}{30}{\micro\metre}) and addressed sequentially and repeatedly at the DMD refresh rate of \SI{30303}{\hertz}. The line rate, given by the DMD rate divided by the number of scan points, was \SIrange{0.8}{3.8}{\kilo\hertz}. Records lasted approximately \SI{180}{\second}. Stacking the successive line scans gives a space--time image (kymograph) in which moving blood cells appear as inclined streaks (Fig. 4c); the blood flow velocity was estimated from their inclination.

% Previous short version:
% \emph{Velocity estimation.} The kymograph was detrended and normalised at each scan point (Fig. 4d) and resampled by spline interpolation onto twice as many points, spaced uniformly in the transit time between neighbouring scan points, which was estimated by cross-correlation. The resampled kymograph was sheared in time using the Fourier shift theorem; at the shear matching the inclination of the streaks, the streaks become vertical (Fig. 4e), which minimises an alignment metric, the mean absolute deviation of the signals across positions (Fig. 4f). A smooth time course of the shear was obtained by relaxation in this metric and converted into velocity, correcting for the sequential acquisition of the scan points\cite{Chaigneau2019} (Fig. 4g,h). The full procedure is given in Supplementary Methods 6. The cardiac frequency was taken as the position of the spectral peak near \SI{10}{\hertz} (Fig. 4i,k), and the distributions in Fig. 4j are kernel density estimates of the velocity over the full record.
\emph{Pre-processing.} The intensity and quality of the foci vary between scan points, and the signal drifts slowly as the tissue moves relative to the field of view. Both effects were removed independently at each scan point by subtracting a moving temporal average and dividing by the moving average of the absolute residual, each over 1,000 line periods (Fig. 4d).

\emph{Resampling of the scan path.} The scan points lie on the square grid of foci, whereas vessels follow curved paths of arbitrary orientation, and only their two-dimensional projection is available. The scan path was smoothed by fitting third-order polynomials to the point coordinates, and its projected length $L$ was taken along the fitted curve. The transit time between neighbouring scan points was estimated from the position of the maximum of the cross-correlation of their signals over the whole record, refined by a parabolic fit. Using spline interpolation, the kymograph was then resampled onto $2N$ points ($N$ being the number of scan points) spaced uniformly in cumulative transit time, so that the streaks become straight. The resampled step is therefore $\delta = L/(2N-1)$. The twofold oversampling reduces the bias of the estimated inclination of fast-moving streaks towards integer-pixel displacements.

\emph{Shear and alignment metric.} The resampled kymograph was sheared in time: the signal at each position was shifted by an amount proportional to its position along the path, using the Fourier shift theorem. At the shear matching the inclination of the streaks, the streaks become vertical, that is, the signals at all positions are aligned in time (Fig. 4e). Alignment was quantified at each time point by the mean absolute deviation of the signals across positions from their mean, which is minimal for matching shear (alignment metric, Fig. 4f). A coarse search (shear between $\pm 5$ line periods per resampled step, in increments of 0.025, averaged over segments of 1,000 lines) identified the range in which the metric dropped by more than 20\% of its maximal reduction. This range was then sampled with 50 shear values for every line of the record. In Fig. 4e,f the shear is given as the total time shift across the scan path in line periods.

\emph{Shear trajectory.} The alignment metric as a function of shear and time was treated as a potential landscape. A smooth trajectory of shear in time was obtained by iterative relaxation under two forces: the negative gradient of the potential with respect to shear, and a smoothness term pulling each point towards the mean of its two temporal neighbours. The trajectory was initialised at the shear minimising the time-averaged potential and updated with an adaptive step until the mean squared net force fell below $10^{-7}$ (estimated shear, Fig. 4f).

\emph{Conversion to velocity.} Because the scan points of a line are acquired one after another, the last point is sampled $(N-1)/N$ of a line period after the first. A cell moving along the scan order therefore needs more time to traverse the path than its streak suggests, and a cell moving against it less, analogous to the correction described for laser-scanning measurements\cite{Chaigneau2019}. With $s$ the estimated shear in line periods per resampled step and $T = N/(\SI{30303}{\hertz})$ the line period, the velocity is
\[
 v = \frac{\delta}{T\,(s+\tau)}, \qquad \tau = \frac{N-1}{N\,(2N-1)},
\]
where $\tau$ is the acquisition delay between neighbouring resampled points in line periods. Positive $v$ denotes flow along the scan order; absolute values are reported (Fig. 4g,h,j).

\emph{Spectral analysis.} The power spectrum of the velocity record (Fig. 4i) was computed as the squared modulus of its Fourier transform. The cardiac frequency was taken as the position of the spectral peak near \SI{10}{\hertz}; Fig. 4k shows the profile of this peak for each record. The distributions in Fig. 4j are kernel density estimates of the velocity over the full record, with the mean marked.

\clearpage
\onecolumn

\makeatletter
\@fleqnfalse
\makeatother

%----------------------------------------------------------------------------------------
%	SUPPLEMENTARY INFORMATION
%----------------------------------------------------------------------------------------

\renewcommand{\thefigure}{S\arabic{figure}}
\renewcommand{\thetable}{S\arabic{table}}
\setcounter{figure}{0}
\setcounter{table}{0}

%\nolinenumbers
\begin{center}
{\Large\bfseries Supplementary Information}\\[0.6em]
\end{center}

\bigskip
\hrule
\bigskip

\section*{Supplementary Note}

\subsection*{Supplementary Note 1: Development of the method}

% Outline (13 Oct 2026). Each stage follows the same template: what was tried, how the problem showed, how it was diagnosed, what fixed it, and the experiments and animals involved. Every stage must have a row in Supplementary Table 1.

The protocol described in the Online Methods was reached through a series of iterations. Before any animal experiments, imaging through the capillary was tested on fluorescent beads embedded in agarose. The first experiments in mice were acute (Stage 1); chronic implantation was developed in the subsequent stages. Each stage was prompted by a failure observed \textit{in vivo}, which was identified, confirmed and eliminated in dedicated experimental sessions before the protocol was changed; in several cases, resolving one issue revealed the next. The stages are described below in chronological order, and the experiments and animals involved are summarised in Supplementary Table 1. All data shown in the main text were acquired with the final protocol.

\subsubsection*{Stage 1: Acute pilot experiments}

In two rounds of acute experiments, the capillary was implanted and imaged in a single session under terminal anaesthesia, after which the mouse was euthanised without regaining consciousness. These experiments confirmed that the probe could reliably enter the capillary, be retracted and be reinserted to find the same structure; full atlas-view imaging was not attempted. The capillary was implanted, and the animal mounted for imaging, with the alignment checked by eye only, which proved insufficient (Stage 3). For calibration, a loose segment of capillary was slipped over the fibre probe, so that the probe was calibrated in whatever shape it had at that moment (Stage 3). The probes were side-view probes with two polished facets, as in our previous work\cite{Stiburek2023}. The implant was filled with water; air bubbles (Stage 4) were very likely already present, but with the alignment issues unresolved, they could not be identified. Two Thy1-GFP M mice were used, with one imaging session each. No data from these experiments are shown.

\subsubsection*{Stage 2: Head plate and implantation protocol}

When the chronic experiments started, the side-view probes were replaced by probes with a single facet polished at \SI{45}{\degree}, which are easier to fabricate and provide a larger field of view; for this application, the field of view was more important than the higher resolution of the probes with two polished facets. To mount the animal reproducibly, we designed a custom head plate that fits both the implantation device and the endomicroscope, so that the same reference is used for implantation and for every imaging session. In the initial implantations, head plate adhesion weakened over time, resulting in a loose bond and occasional detachment. Using a two-component dental cement (Jet Denture Repair Powder and Liquid, Lang Dental) instead of the UV-curable cement (OptiBond FL kit, Kerr) to attach the head plate and cover all exposed bone completely resolved the issue.

Initial capillary implantation also sometimes caused respiratory distress in mice, occasionally progressing to respiratory arrest and death. We hypothesised that these complications resulted from tissue compression during insertion of the rounded capillary tip. Before penetrating the brain, the tip typically indented the pial surface by approximately \SI{800}{\micro\metre}. We suspected that the resulting tissue displacement could affect ventral brain regions involved in respiratory control. To reduce resistance to insertion, we introduced a small incision in the dura mater before capillary implantation. This modification reduced surface indentation and eliminated the observed respiratory complications. Five mice (Thy1-GFP M and C57BL/6) were used to identify and resolve these issues.

\subsubsection*{Stage 3: Alignment of the capillary and the probe}

For probes fabricated from step-index fibres, the geometry of the probe during imaging must match that during calibration, and the probe must enter the capillary along its axis; even a small angular mismatch degrades image quality with increasing imaging depth. This requirement has two parts.

First, the capillary must be implanted with its axis parallel to the axis of the translation stage that inserts it, and the animal must later be mounted under the endomicroscope in the same orientation, so that the probe, suspended from the stage of the endomicroscope, enters the capillary straight. We therefore built an implantation device equipped with two cameras, allowing the capillary to be aligned vertically along the stage axis. During imaging, the mouse was positioned exactly as during surgery using the same head plate and holder. Fibre probes were secured in ferrules with bore diameters closely matched to the fibre diameter, mounted in an ST/PC connector (a49196, Alternativo) to keep the probes vertical.

Second, the free end of a fibre probe glued into its ferrule can bend on its own, and its curvature changes with temperature and humidity. Because the fibre is not resilient to bending, a transmission matrix measured in one shape does not hold in another. Calibration in a loose capillary sleeve (Stage 1) recorded the probe in whatever shape it had at that moment, which then differed from its shape inside the aligned implant. We therefore calibrated the probe inside fixed capillaries glued into grooves milled in the metal frame of the calibration basin (Online Methods; Supplementary Fig. 3), so that it was held in the same straight geometry during calibration and imaging.

With both measures in place, no deterioration in image quality was observed over the full length of the implants. Five imaging sessions in two Thy1-GFP M mice were used to identify these issues and evaluate the modifications.

\subsubsection*{Stage 4: Immersion medium}

The implant was initially filled with deionised water. Its high surface tension prevented the implant from being filled without trapping air bubbles. The bubbles then migrated into the gap between the inner capillary wall and the fibre probe, where they caused severe imaging aberrations. % Previous placeholder: {[}How the bubbles were identified as the cause, e.g. observed under CAM2/CAM3 or inferred from the aberrations.{]} 
The bubbles were visible in the image data themselves: their entry into the field of view could be seen in the recordings, after which they spread over it in the course of the imaging session. The bubbles appeared only intermittently; they were very likely already present in the acute pilot experiments, but could be identified only once the calibration geometry had been fixed (Stage 3). Replacing water with ethanol, whose lower surface tension prevents bubble formation, eliminated the problem (Online Methods). Nineteen imaging sessions in four mice (C57BL/6 and GP5.17) were performed to identify, confirm and eliminate the issue.

\subsubsection*{Stage 5: Cardiac activity motion artefacts}

Cardiac activity induced motion artefacts, mainly in structural imaging. To reduce these artefacts, the scanning sequence was matched to the animal’s heart rate, measured using a pulse oximeter approximately \SI{15}{\minute} after anaesthesia induction. Each imaging plane was divided into rectangular blocks, with dimensions adjusted so that each block was acquired within one cardiac cycle. Corresponding blocks across different focal planes were acquired sequentially over multiple heartbeats, ensuring that spatially adjacent pixels, including those in neighbouring planes, were imaged at the same phase of the cardiac cycle. Further details are given in the Online Methods (Atlas-view imaging). Seven imaging sessions in three Thy1-GFP M mice were used to identify and resolve this issue.

\subsubsection*{Final protocol}

% Previous version: With the final protocol, fourteen mice were implanted, of which six yielded usable imaging {[}reasons for failures, if any: capillary entrance damaged, high motion artefact due to presence of a large vessel around, no response to visual stimulus in GCaMP mice, not optimal sampling in blood flow velocity scans{]}. Structural imaging was performed in 3 mice (ten, four and ten imaging sessions), calcium imaging in 1 mouse (5 imaging sessions) and blood flow measurements in two mice (seven and three imaging sessions). The datasets shown in the main text are from one mouse per modality.
% Previous version: With the final protocol, six mice were implanted. Structural imaging was performed in three mice (ten, four and ten imaging sessions), calcium imaging in one mouse (five imaging sessions) and blood flow measurements in two mice (seven and three imaging sessions). The datasets shown in the main text are from one mouse per modality.\note{Claude}{The previous draft said fourteen final-protocol mice, of which six were usable, and listed reasons for failure (damaged capillary entrance; motion artefacts from a large nearby vessel; no response to visual stimulation in GCaMP mice; suboptimal sampling in blood flow scans). Following Supplementary Table 1 (six final-protocol mice), these failures must belong to the development animals: please confirm, and say in which stage row they are counted (ARRIVE requires exclusions to be reported). Also: the Discussion says calcium imaging in 2 animals, the table and this paragraph say 1.}
With the final protocol, seven mice were implanted, all of which yielded usable imaging; all failures occurred during development and are listed in Supplementary Table 1. Structural imaging was performed in three mice (ten, four and ten imaging sessions), calcium imaging in two mice (five and six imaging sessions) and blood flow measurements in two mice (seven and three imaging sessions). The datasets shown in the main text are from one mouse per modality.

\begin{table}[htb]
\captionsetup{name=Supplementary Table}
\renewcommand{\thetable}{\arabic{table}}
\centering
\small
\caption{\textbf{Development stages and animal use.} Each row lists the imaging sessions and mice used to identify, confirm and eliminate one problem; in the acute experiments (stage 1), each mouse was imaged in a single session; the last rows list the experiments with the final protocol. The total must match the number of implanted mice given in Online Methods (Animals).}
\begin{tabular}{@{}>{\raggedright\arraybackslash}p{0.15\textwidth}>{\raggedright\arraybackslash}p{0.27\textwidth}>{\raggedright\arraybackslash}p{0.25\textwidth}>{\raggedright\arraybackslash}p{0.08\textwidth}>{\raggedright\arraybackslash}p{0.15\textwidth}@{}}
\toprule
Stage & Problem & Diagnosis and solution & Sessions & Mice (line) \\
\midrule
Bench tests & Feasibility of imaging through the capillary & Fluorescent beads in agarose & -- & none \\
1. Acute pilot experiments & Alignment by eye; calibration in a loose capillary sleeve; bubbles not identifiable & Addressed in stages 2--4 & 2 & 2 (Thy1-GFP M) \\
2. Head plate and implantation & Head plate detachment; respiratory distress due to tissue displacement during insertion & Change of the dental cement for fixture to the skull; small incision in the dura & 4 & 5 (Thy1-GFP M, C57BL/6) \\
3. Alignment & Capillary not aligned with the insertion and imaging axis; probe calibrated in a different shape from that during imaging & Two-camera implantation device and a single head-plate holder for implantation and imaging; calibration in fixed capillaries & 5 & 2 (Thy1-GFP M) \\
4. Immersion medium & Intermittent bubbles trapped when filling with water migrated between the probe and the capillary wall, causing severe aberrations & Water replaced by ethanol & 19 & 4 (C57BL/6, GP5.17) \\
5. Motion artefact & Artefacts induced by cardiac activity & Scanning sequence synchronised with the heart rate & 7 & 3 (Thy1-GFP M) \\

\midrule
Final: structure & -- & -- & 24 & 3 (Thy1-GFP M) \\
Final: calcium & -- & -- & 11 & 2 (GP5.17) \\
Final: blood flow & -- & -- & 10 & 2 (C57BL/6J) \\
\midrule
Total & & & 82 & 23 \\
\bottomrule
\end{tabular}
\label{tab:S1}
\end{table}

\clearpage
\section*{Supplementary Figures}

\begin{figure}[htb]
\centering
\includegraphics[width=0.6\textwidth]{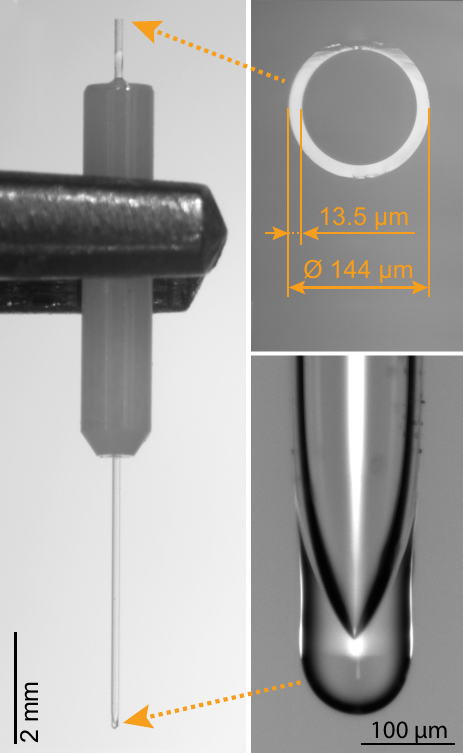}
\caption{\textbf{Capillary implant geometry.} \textbf{a}, The whole implant held with tweezers. \textbf{b}, The implant entrance with dimensions. \textbf{c}, The implant distal end is sealed creating a round tip.}
\label{fig:S1}
\end{figure}

\begin{figure}[htb]
\centering
\includegraphics[width=0.85\textwidth]{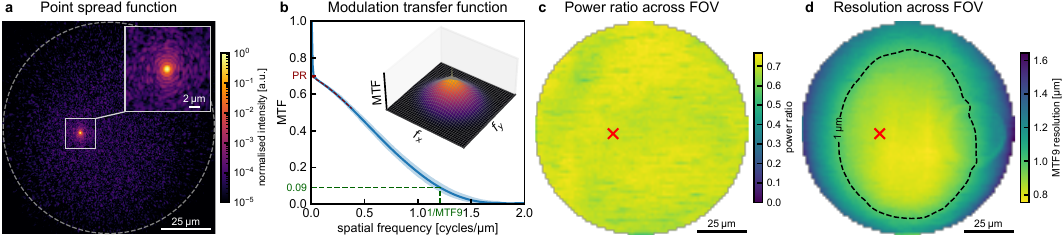}
\caption{\textbf{Imaging performance through the capillary implant.} \textbf{a}, Representative point spread function (PSF) measured through the capillary implant and calibration basin, both filled with water, displayed on a logarithmic intensity scale. The inset shows an enlarged view of the focus. \textbf{b}, Radially averaged modulation transfer function (MTF) calculated from the PSF in \textbf{a}. The blue shaded band indicates $\pm$1~s.d. of the MTF values within each radial spatial-frequency bin. The power ratio (PR) was estimated from the zero-frequency intercept of a fit to the MTF, excluding the zero-frequency component from the fit. Spatial resolution was defined as the reciprocal of the spatial frequency at which the normalised MTF decreased to 0.09. The inset shows a surface plot of the two-dimensional MTF. \textbf{c,d}, Maps of the power ratio (\textbf{c}) and spatial resolution (\textbf{d}) across the field of view, calculated from the corresponding MTFs. Red crosses mark the location of the PSF shown in \textbf{a}. The dashed contour in \textbf{d} delineates the region with spatial resolution better than \SI{1}{\micro\metre}.}
\label{fig:S2}
\end{figure}

\begin{figure}[htb]
\centering
\includegraphics[width=\textwidth,height=0.7\textheight,keepaspectratio]{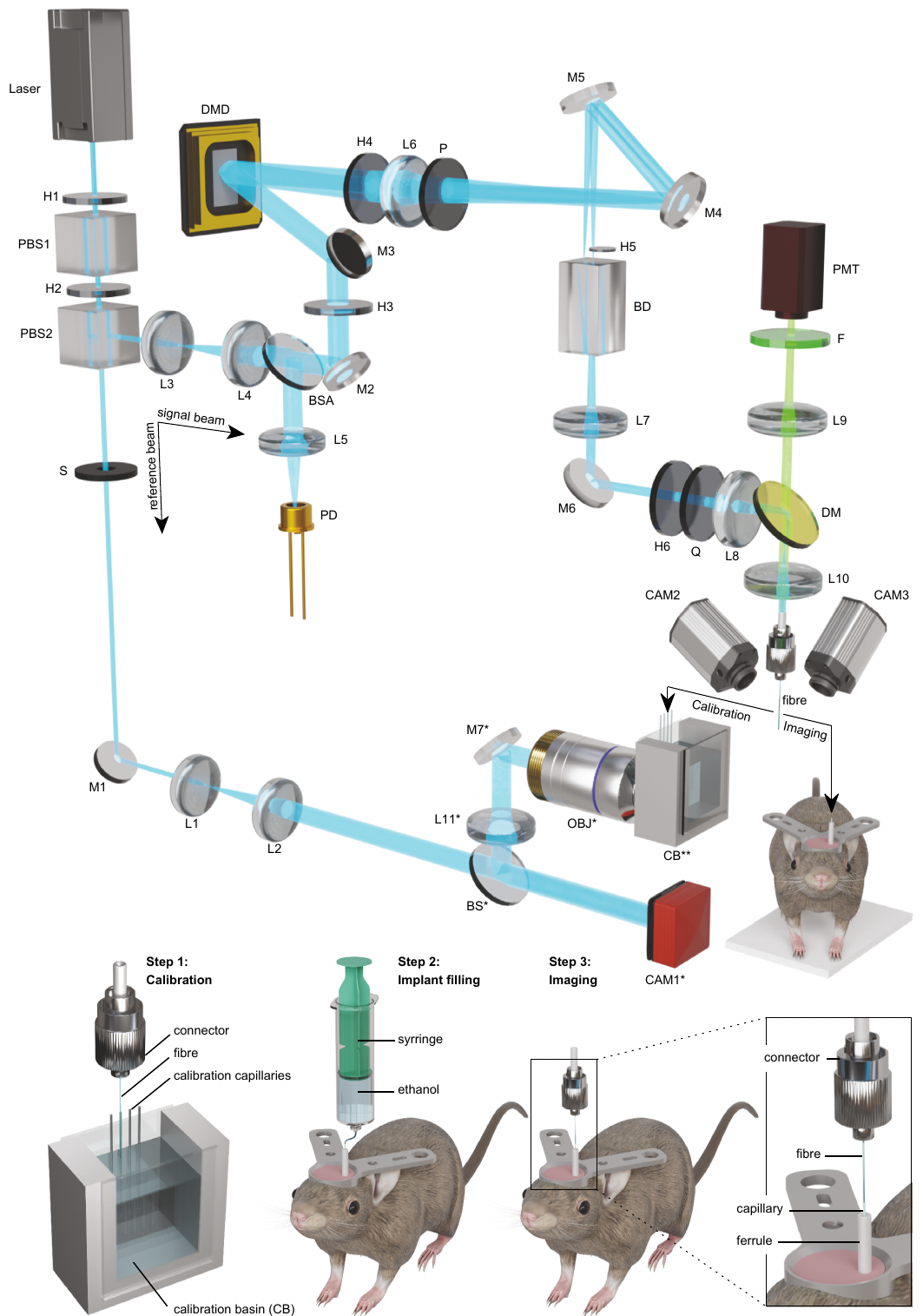}
\caption{\textbf{Experimental geometry.} BD: calcite beam displacer (Thorlabs BD40), BSA: beam sampler (Thorlabs BSF10-A), BS1: beamsplitter plate (Thorlabs BSW10), CAM1: camera (Basler ace acA640-750um), CAM2,3: camera (DinoLite AM4013MTL), CB: calibration basin, DM: dichroic mirror (Chroma T495lpxru), DMD: digital micromirror device (ViALUX V-7001), F: bandpass filter (Chroma ET525/50m), Fibre: multimode fibre-based probe, H1--4: half-wave plate (Edmund Optics \#46-551), H5: half-wave plate (Thorlabs WPH05ME-488), H6: half-wave plate (Edmund Optics \#46-551), L1: lens (Thorlabs C240TMD-A, $f = \SI{8}{\milli\metre}$), L2: lens (Thorlabs AC254-300-A-ML, $f = \SI{300}{\milli\metre}$), L3: lens (Thorlabs C240TMD-A, $f = \SI{8}{\milli\metre}$), L4: lens (Thorlabs AC254-125-A-ML, $f = \SI{125}{\milli\metre}$), L5: lens (Thorlabs AC254-50-A-ML, $f = \SI{50}{\milli\metre}$), L6: lens (Thorlabs AC254-300-A-ML, $f = \SI{300}{\milli\metre}$), L7: lens (Thorlabs AC254-75-A-ML, $f = \SI{75}{\milli\metre}$), L8: lens (Thorlabs AC254-150-A-ML, $f = \SI{150}{\milli\metre}$), L9: lens (Thorlabs AC254-50-A-ML, $f = \SI{50}{\milli\metre}$), L10: lens (Thorlabs C240TMD-A, $f = \SI{8}{\milli\metre}$), L11: lens (Thorlabs AC254-150-A-ML, $f = \SI{150}{\milli\metre}$), Laser: laser (Coherent Sapphire 488 SF NX), M1--7: dielectric mirror (Thorlabs BB1-E02), OBJ: objective (Olympus 20X X-Apo 0.80NA/0.6WD), PBS1,2: polarising beamsplitter (Thorlabs CCM1-PBS251/M), PD: photodetector (Vishay BPW34), PMT: amplified GaAsP photomultiplier tube (Thorlabs PMT2101/M), P: polariser (Thorlabs LPVISA100-MP2), S: shutter (Thorlabs SHB025T), Q: quarter-wave plate (Thorlabs WPMQ10M-488). $\star$~The marked components form the calibration unit. They are rigidly anchored to a single motorised XYZ translation stage composed of Physik Instrumente L-511.2ASD00 and Physik Instrumente L-731.44SD. $\star\star$~The marked calibration basin is housed on a separate XYZ manual translation stage (Thorlabs DT12/M).}
\label{fig:S3}
\end{figure}

\begin{figure}[htb]
\centering
\includegraphics[width=0.85\textwidth]{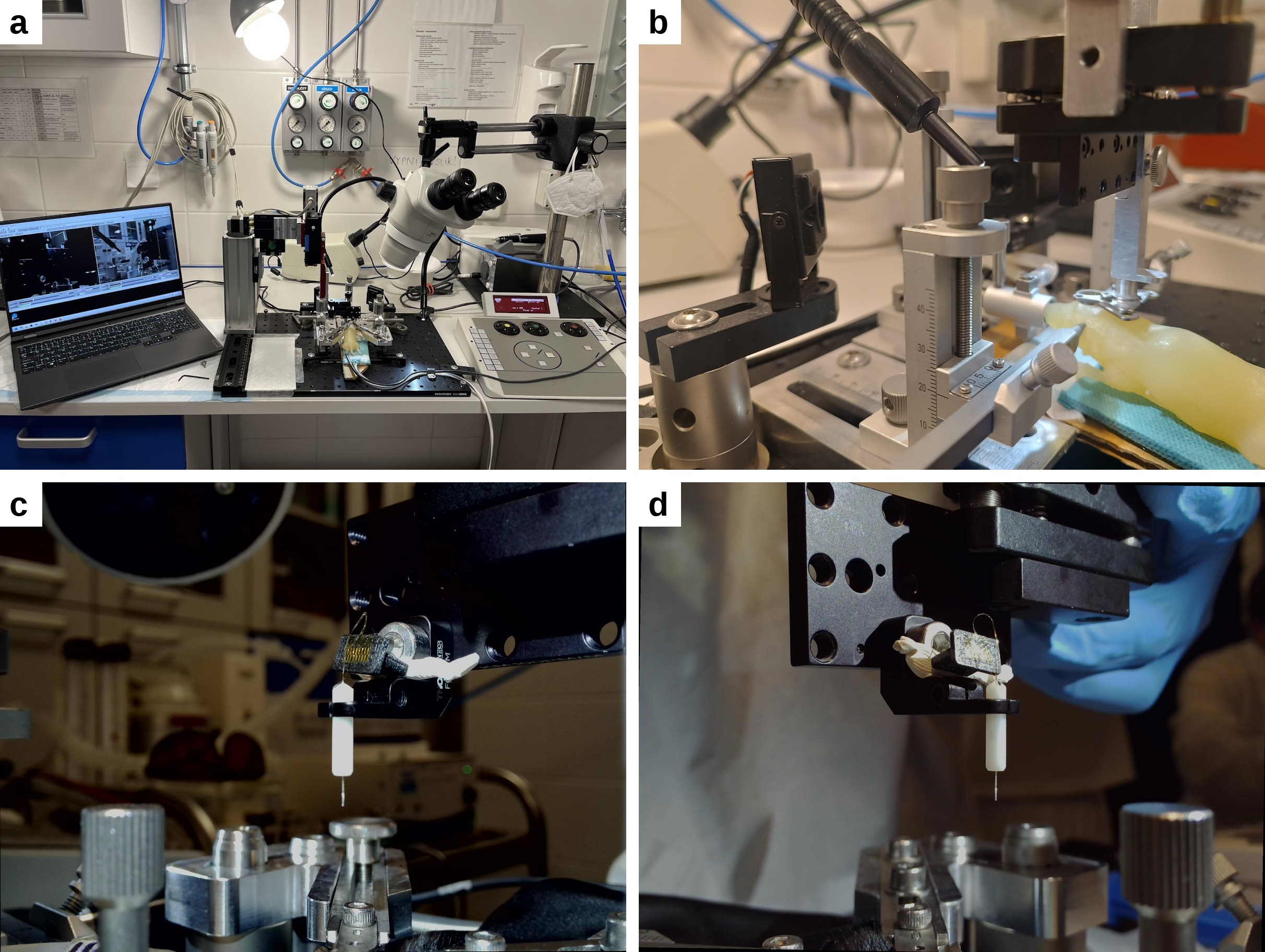}
\caption{\textbf{Implantation set-up.} \textbf{a}, Surgery station with the three-axis manipulator (Unit Mini 25, Luigs \& Neumann) carrying the capillary implant, the head-plate holder, a stereomicroscope and the two alignment cameras, whose views are displayed on the computer (shown with a mouse model). \textbf{b}, Close-up of the head-plate holder, the manipulator arm and one of the alignment cameras, shown with the mouse model. \textbf{c,d}, Views from the two alignment cameras (B03041, Arducam) of an implant in its ferrule, held by the manipulator above the head plate, which is clamped in the holder bars (mouse model); the camera views define the reference for aligning the capillary with the insertion axis.}
\label{fig:S4}
\end{figure}

\clearpage
\section*{Supplementary Videos}

\begin{figure}[!htb]
\centering
\includegraphics[width=\textwidth,height=0.6\textheight,keepaspectratio]{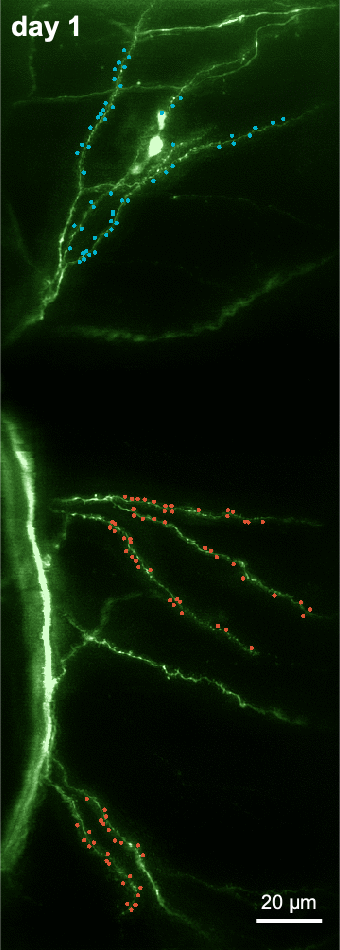}
\caption*{\textbf{Supplementary Video 1: Chronic imaging of a CA1 pyramidal neuron.} Maximum-intensity projection of the processes of a pyramidal neuron with detected basal (blue) and apical (red) spines, animated over all imaging sessions.}
\label{fig:SV1}
\end{figure}

\begin{figure}[!htb]
\centering
\includegraphics[width=0.5\textwidth,height=0.75\textheight,keepaspectratio]{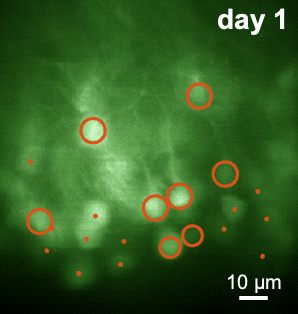}
\caption*{\textbf{Supplementary Video 2: Chronic imaging of dentate gyrus neurons.} All neurons were identified across all imaging sessions.}
\label{fig:SV2}
\end{figure}

\begin{figure}[htb]
\centering
\includegraphics[width=0.6\textwidth,height=0.75\textheight,keepaspectratio]{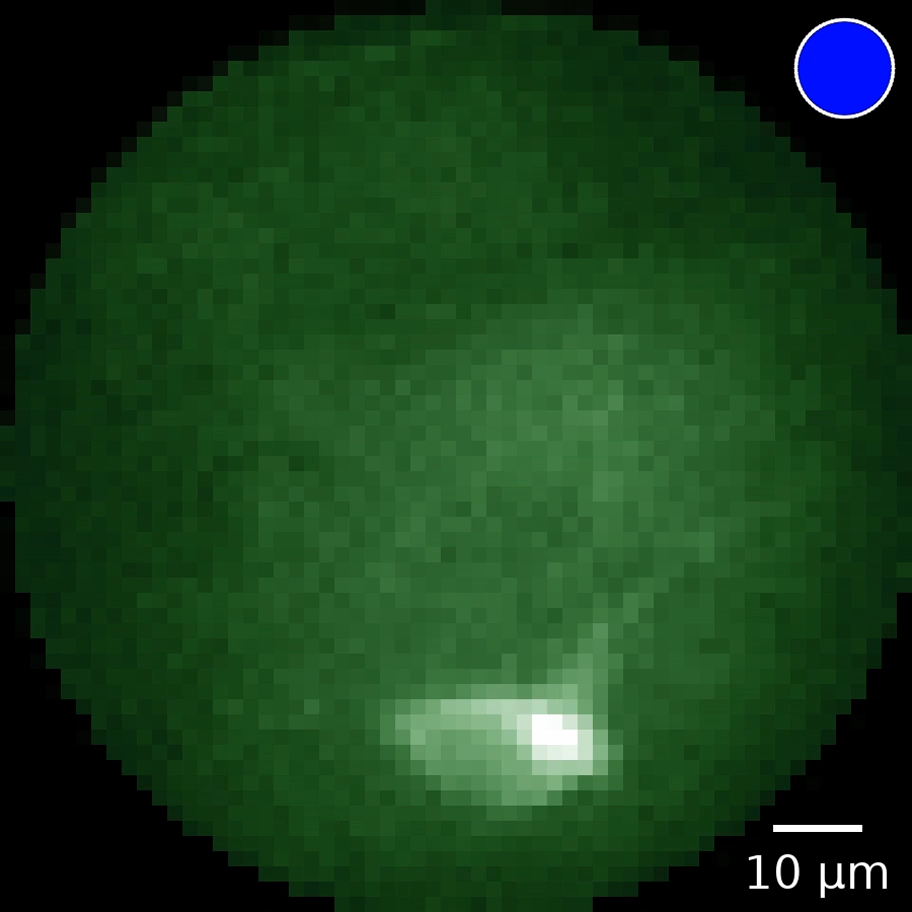}
\caption*{\textbf{Supplementary Video 3: Stimulus-evoked calcium imaging on day 7.} Calcium imaging of Thy1-GCaMP6f-positive pyramidal neurons in cortical layer V on day 7 (related to Fig.~\ref{fig3}) in response to a visual stimulus consisting of blue-light flashes indicated with a blue circle in the upper right corner.}
\label{fig:SV3}
\end{figure}

\begin{figure}[htb]
\centering
\includegraphics[width=\textwidth,height=0.75\textheight,keepaspectratio]{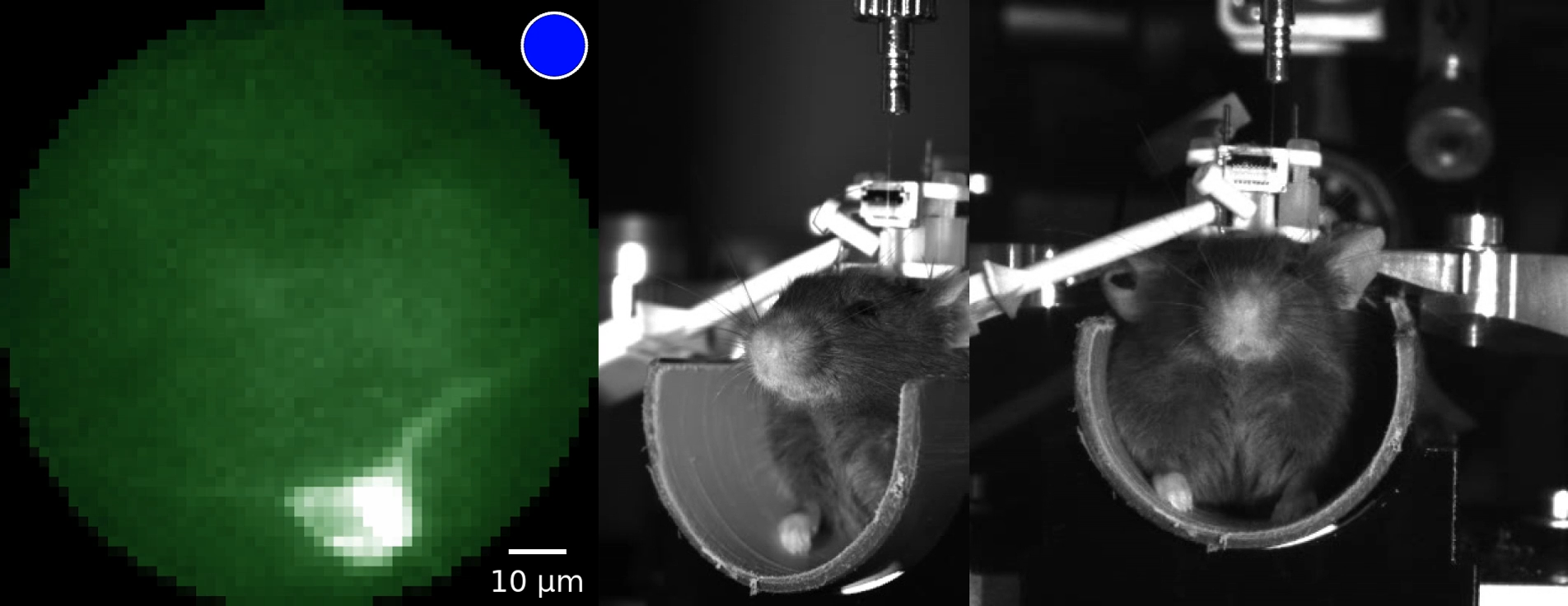}
\caption*{\textbf{Supplementary Video 4: Stimulus-evoked calcium imaging on day 2.} Left, calcium imaging of the same field of view as in Supplementary Video 3, acquired on day 2. The same neurons are responding to the visual stimulus (indicated with a blue circle). Middle, right, simultaneous view of the mouse during the experiment.}
\label{fig:SV4}
\end{figure}

\begin{figure}[htb]
\centering
\includegraphics[width=\textwidth,height=0.75\textheight,keepaspectratio]{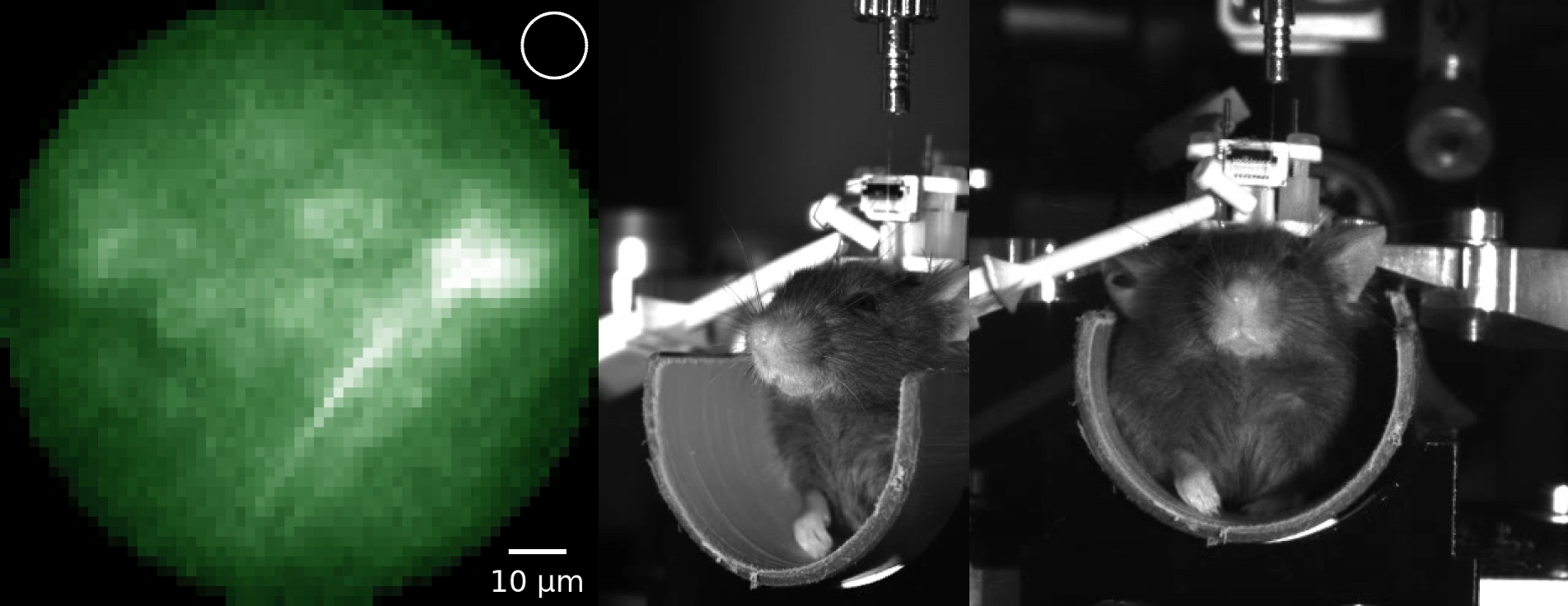}
\caption*{\textbf{Supplementary Video 5: Calcium activity in hippocampal neurons on day 2.} Left, calcium imaging in the hippocampus during the same imaging session as in Supplementary Video 4. While cortical neurons respond to the visual stimulus, hippocampal neurons show rather spontaneous activity. Middle, right, simultaneous view of the mouse during the experiment.}
\label{fig:SV5}
\end{figure}

\begin{figure}[htb]
\centering
\includegraphics[width=0.6\textwidth,height=0.75\textheight,keepaspectratio]{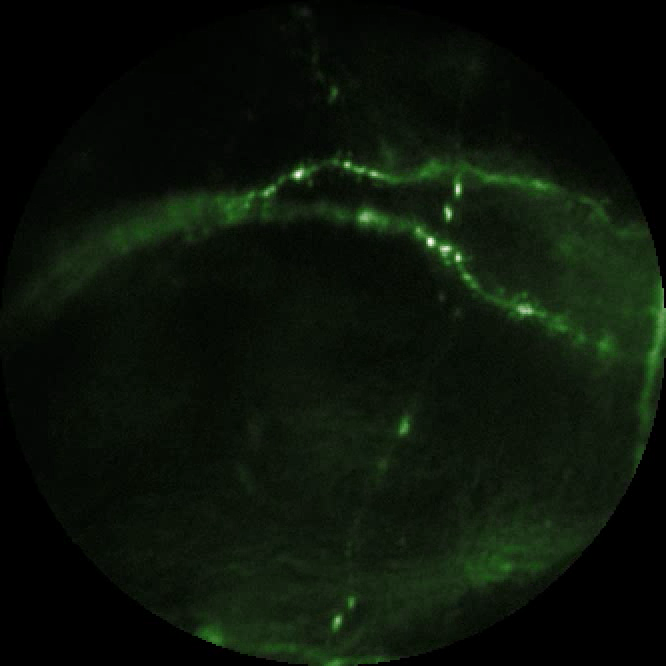}
\caption*{\textbf{Supplementary Video 6: Circumferential imaging by rotation about the capillary axis.} The probe was advanced along the capillary, and at multiple axial positions the sample was rotated about the capillary axis to acquire a \SI{360}{\degree} panoramic view.}
\label{fig:SV6}
\end{figure}

\end{document}